\def\arxiv{}
\ifdefined\arxiv
    \def\acmopts{sigplan,authorversion}
\else
    \def\acmopts{sigplan,screen}
\fi

\documentclass[\acmopts]{acmart}
\usepackage[T1]{fontenc}

\usepackage{subcaption}

\setcopyright{acmlicensed}
\acmPrice{15.00}
\acmDOI{10.1145/3573105.3575673}
\acmYear{2023}
\copyrightyear{2023}
\acmSubmissionID{poplws23cppmain-p20-p}
\acmISBN{979-8-4007-0026-2/23/01}
\acmConference[CPP '23]{Proceedings of the 12th ACM SIGPLAN International Conference on Certified Programs and Proofs}{January 16--17, 2023}{Boston, MA, USA}
\acmBooktitle{Proceedings of the 12th ACM SIGPLAN International Conference on Certified Programs and Proofs (CPP '23), January 16--17, 2023, Boston, MA, USA}
\received{2022-09-21}
\received[accepted]{2022-11-21}

\title{Verifying Term Graph Optimizations using Isabelle/HOL}

\author{Brae J. Webb}
\orcid{0000-0003-3579-0244}
\affiliation{
  \institution{The University of Queensland}
  \city{Brisbane}
  \country{Australia}
}
\email{b.webb@uq.edu.au}

\author{Ian J. Hayes}
\orcid{0000-0003-3649-392X}
\affiliation{
  \institution{The University of Queensland}
  \city{Brisbane}
  \country{Australia}
}
\email{ian.hayes@uq.edu.au}

\author{Mark Utting}
\orcid{0000-0003-3134-6306}
\affiliation{
  \institution{The University of Queensland}
  \city{Brisbane}
  \country{Australia}
}
\email{m.utting@uq.edu.au}



\usepackage{isabelle,isabellesym}
\usepackage{mathpartir}
\isabellestyle{it}

\usepackage{paper}
\Snip{algebraic-laws}{%
\isamarkuptrue%
\begin{isamarkuptext}%
\begin{align}
\isa{x\ {\isacharminus}{\kern0pt}\ x\ {\isacharequal}{\kern0pt}\ {\isadigit{0}}}\\
\isa{x\ {\isacharminus}{\kern0pt}\ {\isacharparenleft}{\kern0pt}x\ {\isacharminus}{\kern0pt}\ y{\isacharparenright}{\kern0pt}\ {\isacharequal}{\kern0pt}\ y}
\end{align}%
\end{isamarkuptext}\isamarkuptrue%
}%
\Snip{algebraic-laws-values}{%
\isamarkuptrue%
\begin{align}
\isa{{\isasymforall}x\ {\isacharcolon}{\kern0pt}{\isacharcolon}{\kern0pt}\ {\isacharprime}{\kern0pt}a\ word{\isachardot}{\kern0pt}\ x\ {\isacharminus}{\kern0pt}\ x\ {\isacharequal}{\kern0pt}\ {\isacharparenleft}{\kern0pt}{\isadigit{0}}\ {\isacharcolon}{\kern0pt}{\isacharcolon}{\kern0pt}\ {\isacharprime}{\kern0pt}a\ word{\isacharparenright}{\kern0pt}} \label{prop-v-minus-v}\\
\isa{{\isasymforall}{\isacharparenleft}{\kern0pt}x{\isacharcolon}{\kern0pt}{\isacharcolon}{\kern0pt}{\isacharprime}{\kern0pt}a\ word{\isacharparenright}{\kern0pt}\ y\ {\isacharcolon}{\kern0pt}{\isacharcolon}{\kern0pt}\ {\isacharprime}{\kern0pt}a\ word{\isachardot}{\kern0pt}\ x\ {\isacharminus}{\kern0pt}\ {\isacharparenleft}{\kern0pt}x\ {\isacharminus}{\kern0pt}\ y{\isacharparenright}{\kern0pt}\ {\isacharequal}{\kern0pt}\ y} \label{prop-redundant-sub}
\end{align}
}%
\Snip{algebraic-laws-expressions}{%
\isamarkuptrue%
\begin{align}
\isa{x\ {\isacharminus}{\kern0pt}\ x\ {\isasymlongmapsto}\ {\isadigit{0}}} \label{prop-MinusSame} \\
\isa{x\ {\isacharminus}{\kern0pt}\ {\isacharparenleft}{\kern0pt}x\ {\isacharminus}{\kern0pt}\ y{\isacharparenright}{\kern0pt}\ {\isasymlongmapsto}\ y}
\end{align}
}%
\Snip{sub-same-32}{%
\isamarkuptrue%
\isacommand{optimization}\isamarkupfalse%
\ SubIdentity{\isacharcolon}{\kern0pt}\isanewline
\ \ {\isachardoublequoteopen}x\ {\isacharminus}{\kern0pt}\ x\ {\isasymlongmapsto}\ ConstantExpr\ {\isacharparenleft}{\kern0pt}IntVal\ b\ {\isadigit{0}}{\isacharparenright}{\kern0pt}\isanewline
\ \ \ \ \ when\ {\isacharparenleft}{\kern0pt}{\isacharparenleft}{\kern0pt}stamp{\isacharunderscore}{\kern0pt}expr\ exp{\isacharbrackleft}{\kern0pt}x\ {\isacharminus}{\kern0pt}\ x{\isacharbrackright}{\kern0pt}\ {\isacharequal}{\kern0pt}\ IntegerStamp\ b\ lo\ hi{\isacharparenright}{\kern0pt}\ {\isasymand}\ wf{\isacharunderscore}{\kern0pt}stamp\ exp{\isacharbrackleft}{\kern0pt}x\ {\isacharminus}{\kern0pt}\ x{\isacharbrackright}{\kern0pt}{\isacharparenright}{\kern0pt}{\isachardoublequoteclose}%
}%
\Snip{RedundantSubtract}{%
\isamarkuptrue%
\isacommand{optimization}\isamarkupfalse%
\ RedundantSubtract{\isacharcolon}{\kern0pt}\isanewline
\ \ {\isachardoublequoteopen}x\ {\isacharminus}{\kern0pt}\ {\isacharparenleft}{\kern0pt}x\ {\isacharminus}{\kern0pt}\ y{\isacharparenright}{\kern0pt}\ {\isasymlongmapsto}\ y{\isachardoublequoteclose}%
}%
\Snip{ast-example}{%
\isamarkuptrue%
\begin{isamarkuptext}%
\begin{isabelle}%
BinaryExpr\ BinAdd\isanewline
\isaindent{\ }{\isacharparenleft}{\kern0pt}BinaryExpr\ BinMul\ x\ x{\isacharparenright}{\kern0pt}\isanewline
\isaindent{\ }{\isacharparenleft}{\kern0pt}BinaryExpr\ BinMul\ x\ x{\isacharparenright}{\kern0pt}%
\end{isabelle}%
\end{isamarkuptext}\isamarkuptrue%
}%
\Snip{abstract-syntax-tree}{%
\isamarkuptrue%
\begin{isamarkuptext}%
\begin{isabelle}%
\isacommand{datatype}\ IRExpr\ {\isacharequal}{\kern0pt}\isanewline
\isaindent{\ \ }UnaryExpr\ IRUnaryOp\ IRExpr\isanewline
\isaindent{\ \ }{\isacharbar}{\kern0pt}\ BinaryExpr\ IRBinaryOp\ IRExpr\ IRExpr\isanewline
\isaindent{\ \ }{\isacharbar}{\kern0pt}\ ConditionalExpr\ IRExpr\ IRExpr\ IRExpr\isanewline
\isaindent{\ \ }{\isacharbar}{\kern0pt}\ ParameterExpr\ nat\ Stamp\isanewline
\isaindent{\ \ }{\isacharbar}{\kern0pt}\ LeafExpr\ nat\ Stamp\isanewline
\isaindent{\ \ }{\isacharbar}{\kern0pt}\ ConstantExpr\ Value\isanewline
\isaindent{\ \ }{\isacharbar}{\kern0pt}\ ConstantVar\ {\isacharparenleft}{\kern0pt}char\ list{\isacharparenright}{\kern0pt}\isanewline
\isaindent{\ \ }{\isacharbar}{\kern0pt}\ VariableExpr\ {\isacharparenleft}{\kern0pt}char\ list{\isacharparenright}{\kern0pt}\ Stamp%
\end{isabelle}%
\end{isamarkuptext}\isamarkuptrue%
}%
\Snip{value}{%
\isamarkuptrue%
\begin{isamarkuptext}%
\begin{isabelle}%
\isacommand{datatype}\ Value\ {\isacharequal}{\kern0pt}\ UndefVal\isanewline
\isaindent{\ \ }{\isacharbar}{\kern0pt}\ IntVal\ nat\ {\isacharparenleft}{\kern0pt}{\isadigit{6}}{\isadigit{4}}\ word{\isacharparenright}{\kern0pt}\isanewline
\isaindent{\ \ }{\isacharbar}{\kern0pt}\ ObjRef\ {\isacharparenleft}{\kern0pt}nat\ option{\isacharparenright}{\kern0pt}\isanewline
\isaindent{\ \ }{\isacharbar}{\kern0pt}\ ObjStr\ {\isacharparenleft}{\kern0pt}char\ list{\isacharparenright}{\kern0pt}%
\end{isabelle}%
\end{isamarkuptext}\isamarkuptrue%
}%
\Snip{eval}{%
\isamarkuptrue%
\begin{isamarkuptext}%
\isa{unary{\isacharunderscore}{\kern0pt}eval\ {\isacharcolon}{\kern0pt}{\isacharcolon}{\kern0pt}\ IRUnaryOp\ {\isasymRightarrow}\ Value\ {\isasymRightarrow}\ Value}\\
\isa{bin{\isacharunderscore}{\kern0pt}eval\ {\isacharcolon}{\kern0pt}{\isacharcolon}{\kern0pt}\ IRBinaryOp\ {\isasymRightarrow}\ Value\ {\isasymRightarrow}\ Value\ {\isasymRightarrow}\ Value}%
\end{isamarkuptext}\isamarkuptrue%
}%
\Snip{tree-semantics}{%
\isamarkuptrue%
\begin{isamarkuptext}%
\induct{\isa{\mbox{}\inferrule{\mbox{{\isacharbrackleft}{\kern0pt}m{\isacharcomma}{\kern0pt}p{\isacharbrackright}{\kern0pt}\ {\isasymturnstile}\ xe\ {\isasymmapsto}\ x}\\\ \mbox{result\ {\isacharequal}{\kern0pt}\ unary{\isacharunderscore}{\kern0pt}eval\ op\ x}\\\ \mbox{result\ {\isasymnoteq}\ UndefVal}}{\mbox{{\isacharbrackleft}{\kern0pt}m{\isacharcomma}{\kern0pt}p{\isacharbrackright}{\kern0pt}\ {\isasymturnstile}\ UnaryExpr\ op\ xe\ {\isasymmapsto}\ result}}}}{semantics:unary}
\induct{\isa{\mbox{}\inferrule{\mbox{{\isacharbrackleft}{\kern0pt}m{\isacharcomma}{\kern0pt}p{\isacharbrackright}{\kern0pt}\ {\isasymturnstile}\ xe\ {\isasymmapsto}\ x}\\\ \mbox{{\isacharbrackleft}{\kern0pt}m{\isacharcomma}{\kern0pt}p{\isacharbrackright}{\kern0pt}\ {\isasymturnstile}\ ye\ {\isasymmapsto}\ y}\\\ \mbox{result\ {\isacharequal}{\kern0pt}\ bin{\isacharunderscore}{\kern0pt}eval\ op\ x\ y}\\\ \mbox{result\ {\isasymnoteq}\ UndefVal}}{\mbox{{\isacharbrackleft}{\kern0pt}m{\isacharcomma}{\kern0pt}p{\isacharbrackright}{\kern0pt}\ {\isasymturnstile}\ BinaryExpr\ op\ xe\ ye\ {\isasymmapsto}\ result}}}}{semantics:binary}
\induct{\isa{\mbox{}\inferrule{\mbox{{\isacharbrackleft}{\kern0pt}m{\isacharcomma}{\kern0pt}p{\isacharbrackright}{\kern0pt}\ {\isasymturnstile}\ ce\ {\isasymmapsto}\ cond}\\\ \mbox{cond\ {\isasymnoteq}\ UndefVal}\\\ \mbox{branch\ {\isacharequal}{\kern0pt}\ {\isacharparenleft}{\kern0pt}\textsf{if}\ bool-of\ cond\ \textsf{then}\ te\ \textsf{else}\ fe{\isacharparenright}{\kern0pt}}\\\ \mbox{{\isacharbrackleft}{\kern0pt}m{\isacharcomma}{\kern0pt}p{\isacharbrackright}{\kern0pt}\ {\isasymturnstile}\ branch\ {\isasymmapsto}\ result}\\\ \mbox{result\ {\isasymnoteq}\ UndefVal}}{\mbox{{\isacharbrackleft}{\kern0pt}m{\isacharcomma}{\kern0pt}p{\isacharbrackright}{\kern0pt}\ {\isasymturnstile}\ ConditionalExpr\ ce\ te\ fe\ {\isasymmapsto}\ result}}}}{semantics:conditional}
\induct{\isa{\mbox{}\inferrule{\mbox{wf{\isacharunderscore}{\kern0pt}value\ c}}{\mbox{{\isacharbrackleft}{\kern0pt}m{\isacharcomma}{\kern0pt}p{\isacharbrackright}{\kern0pt}\ {\isasymturnstile}\ ConstantExpr\ c\ {\isasymmapsto}\ c}}}}{semantics:constant}
\induct{\isa{\mbox{}\inferrule{\mbox{i\ {\isacharless}{\kern0pt}\ {\isacharbar}{\kern0pt}p{\isacharbar}{\kern0pt}}\\\ \mbox{p\ensuremath{_{[\mathit{i}]}}\ {\isasymin}\ s}}{\mbox{{\isacharbrackleft}{\kern0pt}m{\isacharcomma}{\kern0pt}p{\isacharbrackright}{\kern0pt}\ {\isasymturnstile}\ ParameterExpr\ i\ s\ {\isasymmapsto}\ p\ensuremath{_{[\mathit{i}]}}}}}}{semantics:parameter}
\induct{\isa{\mbox{}\inferrule{\mbox{val\ {\isacharequal}{\kern0pt}\ m\ n}\\\ \mbox{val\ {\isasymin}\ s}}{\mbox{{\isacharbrackleft}{\kern0pt}m{\isacharcomma}{\kern0pt}p{\isacharbrackright}{\kern0pt}\ {\isasymturnstile}\ LeafExpr\ n\ s\ {\isasymmapsto}\ val}}}}{semantics:leaf}%
\end{isamarkuptext}\isamarkuptrue%
}%
\Snip{tree-evaluation-deterministic}{%
\isamarkuptrue%
\begin{isamarkuptext}%
\begin{isabelle}%
{\isacharbrackleft}{\kern0pt}m{\isacharcomma}{\kern0pt}p{\isacharbrackright}{\kern0pt}\ {\isasymturnstile}\ e\ {\isasymmapsto}\ v\isactrlsub {\isadigit{1}}\ {\isasymand}\ {\isacharbrackleft}{\kern0pt}m{\isacharcomma}{\kern0pt}p{\isacharbrackright}{\kern0pt}\ {\isasymturnstile}\ e\ {\isasymmapsto}\ v\isactrlsub {\isadigit{2}}\ {\isasymLongrightarrow}\ v\isactrlsub {\isadigit{1}}\ {\isacharequal}{\kern0pt}\ v\isactrlsub {\isadigit{2}}%
\end{isabelle}%
\end{isamarkuptext}\isamarkuptrue%
}%
\Snip{expression-refinement}{%
\isamarkuptrue%
\begin{isamarkuptext}%
\isa{e\isactrlsub {\isadigit{1}}\ {\isasymsqsupseteq}\ e\isactrlsub {\isadigit{2}}\ {\isacharequal}{\kern0pt}\ {\isacharparenleft}{\kern0pt}{\isasymforall}m\ p\ v{\isachardot}{\kern0pt}\ {\isacharbrackleft}{\kern0pt}m{\isacharcomma}{\kern0pt}p{\isacharbrackright}{\kern0pt}\ {\isasymturnstile}\ e\isactrlsub {\isadigit{1}}\ {\isasymmapsto}\ v\ {\isasymlongrightarrow}\ {\isacharbrackleft}{\kern0pt}m{\isacharcomma}{\kern0pt}p{\isacharbrackright}{\kern0pt}\ {\isasymturnstile}\ e\isactrlsub {\isadigit{2}}\ {\isasymmapsto}\ v{\isacharparenright}{\kern0pt}}%
\end{isamarkuptext}\isamarkuptrue%
}%
\Snip{InverseLeftSub}{%
\isamarkuptrue%
\isacommand{optimization}\isamarkupfalse%
\ InverseLeftSub{\isacharcolon}{\kern0pt}\isanewline
\ \ {\isachardoublequoteopen}{\isacharparenleft}{\kern0pt}x\ {\isacharminus}{\kern0pt}\ y{\isacharparenright}{\kern0pt}\ {\isacharplus}{\kern0pt}\ y\ {\isasymlongmapsto}\ x{\isachardoublequoteclose}%
}%
\Snip{InverseLeftSubObligation}{%
\isamarkuptrue%
\begin{isamarkuptext}%
\begin{isabelle}%
\ {\isadigit{1}}{\isachardot}{\kern0pt}\ trm(x)\ {\isacharless}{\kern0pt}\ trm(BinaryExpr\ BinAdd\ {\isacharparenleft}{\kern0pt}BinaryExpr\ BinSub\ x\ y{\isacharparenright}{\kern0pt}\ y)\isanewline
\ {\isadigit{2}}{\isachardot}{\kern0pt}\ BinaryExpr\ BinAdd\ {\isacharparenleft}{\kern0pt}BinaryExpr\ BinSub\ x\ y{\isacharparenright}{\kern0pt}\ y\ {\isasymsqsupseteq}\ x%
\end{isabelle}%
\end{isamarkuptext}\isamarkuptrue%
}%
\Snip{InverseRightSub}{%
\isamarkuptrue%
\isacommand{optimization}\isamarkupfalse%
\ InverseRightSub{\isacharcolon}{\kern0pt}\ {\isachardoublequoteopen}y\ {\isacharplus}{\kern0pt}\ {\isacharparenleft}{\kern0pt}x\ {\isacharminus}{\kern0pt}\ y{\isacharparenright}{\kern0pt}\ {\isasymlongmapsto}\ x{\isachardoublequoteclose}%
}%
\Snip{InverseRightSubObligation}{%
\isamarkuptrue%
\begin{isamarkuptext}%
\begin{isabelle}%
\ {\isadigit{1}}{\isachardot}{\kern0pt}\ trm(x)\ {\isacharless}{\kern0pt}\ trm(BinaryExpr\ BinAdd\ y\ {\isacharparenleft}{\kern0pt}BinaryExpr\ BinSub\ x\ y{\isacharparenright}{\kern0pt})\isanewline
\ {\isadigit{2}}{\isachardot}{\kern0pt}\ BinaryExpr\ BinAdd\ y\ {\isacharparenleft}{\kern0pt}BinaryExpr\ BinSub\ x\ y{\isacharparenright}{\kern0pt}\ {\isasymsqsupseteq}\ x%
\end{isabelle}%
\end{isamarkuptext}\isamarkuptrue%
}%
\Snip{expression-refinement-monotone}{%
\isamarkuptrue%
\begin{isamarkuptext}%
\begin{isabelle}%
x\ {\isasymsqsupseteq}\ x{\isacharprime}{\kern0pt}\ {\isasymLongrightarrow}\ UnaryExpr\ op\ x\ {\isasymsqsupseteq}\ UnaryExpr\ op\ x{\isacharprime}{\kern0pt}%
\end{isabelle}
\begin{isabelle}%
x\ {\isasymsqsupseteq}\ x{\isacharprime}{\kern0pt}\ {\isasymand}\ y\ {\isasymsqsupseteq}\ y{\isacharprime}{\kern0pt}\ {\isasymLongrightarrow}\ BinaryExpr\ op\ x\ y\ {\isasymsqsupseteq}\ BinaryExpr\ op\ x{\isacharprime}{\kern0pt}\ y{\isacharprime}{\kern0pt}%
\end{isabelle}
\begin{isabelle}%
c\ {\isasymsqsupseteq}\ c{\isacharprime}{\kern0pt}\ {\isasymand}\ t\ {\isasymsqsupseteq}\ t{\isacharprime}{\kern0pt}\ {\isasymand}\ f\ {\isasymsqsupseteq}\ f{\isacharprime}{\kern0pt}\ {\isasymLongrightarrow}\isanewline
ConditionalExpr\ c\ t\ f\ {\isasymsqsupseteq}\ ConditionalExpr\ c{\isacharprime}{\kern0pt}\ t{\isacharprime}{\kern0pt}\ f{\isacharprime}{\kern0pt}%
\end{isabelle}%
\end{isamarkuptext}\isamarkuptrue%
}%
\Snip{BinaryFoldConstant}{%
\isamarkuptrue%
\isacommand{optimization}\isamarkupfalse%
\ BinaryFoldConstant{\isacharcolon}{\kern0pt}\ {\isachardoublequoteopen}BinaryExpr\ op\ {\isacharparenleft}{\kern0pt}const\ v{\isadigit{1}}{\isacharparenright}{\kern0pt}\ {\isacharparenleft}{\kern0pt}const\ v{\isadigit{2}}{\isacharparenright}{\kern0pt}\ {\isasymlongmapsto}\ ConstantExpr\ {\isacharparenleft}{\kern0pt}bin{\isacharunderscore}{\kern0pt}eval\ op\ v{\isadigit{1}}\ v{\isadigit{2}}{\isacharparenright}{\kern0pt}{\isachardoublequoteclose}%
}%
\Snip{BinaryFoldConstantObligation}{%
\isamarkuptrue%
\begin{isamarkuptext}%
\begin{isabelle}%
\ {\isadigit{1}}{\isachardot}{\kern0pt}\ trm(ConstantExpr\ {\isacharparenleft}{\kern0pt}bin{\isacharunderscore}{\kern0pt}eval\ op\ v{\isadigit{1}}\ v{\isadigit{2}}{\isacharparenright}{\kern0pt})\isanewline
\isaindent{\ {\isadigit{1}}{\isachardot}{\kern0pt}\ }{\isacharless}{\kern0pt}\ trm(BinaryExpr\ op\ {\isacharparenleft}{\kern0pt}ConstantExpr\ v{\isadigit{1}}{\isacharparenright}{\kern0pt}\ {\isacharparenleft}{\kern0pt}ConstantExpr\ v{\isadigit{2}}{\isacharparenright}{\kern0pt})\isanewline
\ {\isadigit{2}}{\isachardot}{\kern0pt}\ BinaryExpr\ op\ {\isacharparenleft}{\kern0pt}ConstantExpr\ v{\isadigit{1}}{\isacharparenright}{\kern0pt}\ {\isacharparenleft}{\kern0pt}ConstantExpr\ v{\isadigit{2}}{\isacharparenright}{\kern0pt}\ {\isasymsqsupseteq}\isanewline
\isaindent{\ {\isadigit{2}}{\isachardot}{\kern0pt}\ }ConstantExpr\ {\isacharparenleft}{\kern0pt}bin{\isacharunderscore}{\kern0pt}eval\ op\ v{\isadigit{1}}\ v{\isadigit{2}}{\isacharparenright}{\kern0pt}%
\end{isabelle}%
\end{isamarkuptext}\isamarkuptrue%
}%
\Snip{AddCommuteConstantRight}{%
\isamarkuptrue%
\isacommand{optimization}\isamarkupfalse%
\ AddCommuteConstantRight{\isacharcolon}{\kern0pt}\isanewline
\ \ {\isachardoublequoteopen}{\isacharparenleft}{\kern0pt}const\ v{\isacharparenright}{\kern0pt}\ {\isacharplus}{\kern0pt}\ y\ {\isasymlongmapsto}\ y\ {\isacharplus}{\kern0pt}\ {\isacharparenleft}{\kern0pt}const\ v{\isacharparenright}{\kern0pt}\ when\ {\isasymnot}{\isacharparenleft}{\kern0pt}is{\isacharunderscore}{\kern0pt}ConstantExpr\ y{\isacharparenright}{\kern0pt}{\isachardoublequoteclose}%
}%
\Snip{AddCommuteConstantRightObligation}{%
\isamarkuptrue%
\begin{isamarkuptext}%
\begin{isabelle}%
\ {\isadigit{1}}{\isachardot}{\kern0pt}\ {\isasymnot}\ is{\isacharunderscore}{\kern0pt}ConstantExpr\ y\ {\isasymlongrightarrow}\isanewline
\isaindent{\ {\isadigit{1}}{\isachardot}{\kern0pt}\ }trm(BinaryExpr\ BinAdd\ y\ {\isacharparenleft}{\kern0pt}ConstantExpr\ v{\isacharparenright}{\kern0pt})\isanewline
\isaindent{\ {\isadigit{1}}{\isachardot}{\kern0pt}\ }{\isacharless}{\kern0pt}\ trm(BinaryExpr\ BinAdd\ {\isacharparenleft}{\kern0pt}ConstantExpr\ v{\isacharparenright}{\kern0pt}\ y)\isanewline
\ {\isadigit{2}}{\isachardot}{\kern0pt}\ {\isasymnot}\ is{\isacharunderscore}{\kern0pt}ConstantExpr\ y\ {\isasymlongrightarrow}\isanewline
\isaindent{\ {\isadigit{2}}{\isachardot}{\kern0pt}\ }BinaryExpr\ BinAdd\ {\isacharparenleft}{\kern0pt}ConstantExpr\ v{\isacharparenright}{\kern0pt}\ y\ {\isasymsqsupseteq}\isanewline
\isaindent{\ {\isadigit{2}}{\isachardot}{\kern0pt}\ }BinaryExpr\ BinAdd\ y\ {\isacharparenleft}{\kern0pt}ConstantExpr\ v{\isacharparenright}{\kern0pt}%
\end{isabelle}%
\end{isamarkuptext}\isamarkuptrue%
}%
\Snip{AddNeutral}{%
\isamarkuptrue%
\isacommand{optimization}\isamarkupfalse%
\ AddNeutral{\isacharcolon}{\kern0pt}\ {\isachardoublequoteopen}x\ {\isacharplus}{\kern0pt}\ {\isacharparenleft}{\kern0pt}const\ {\isacharparenleft}{\kern0pt}IntVal\ {\isadigit{3}}{\isadigit{2}}\ {\isadigit{0}}{\isacharparenright}{\kern0pt}{\isacharparenright}{\kern0pt}\ {\isasymlongmapsto}\ x{\isachardoublequoteclose}%
}%
\Snip{AddNeutralObligation}{%
\isamarkuptrue%
\begin{isamarkuptext}%
\begin{isabelle}%
\ {\isadigit{1}}{\isachardot}{\kern0pt}\ trm(x)\ {\isacharless}{\kern0pt}\ trm(BinaryExpr\ BinAdd\ x\ {\isacharparenleft}{\kern0pt}ConstantExpr\ {\isacharparenleft}{\kern0pt}IntVal\ {\isadigit{3}}{\isadigit{2}}\ {\isadigit{0}}{\isacharparenright}{\kern0pt}{\isacharparenright}{\kern0pt})\isanewline
\ {\isadigit{2}}{\isachardot}{\kern0pt}\ BinaryExpr\ BinAdd\ x\ {\isacharparenleft}{\kern0pt}ConstantExpr\ {\isacharparenleft}{\kern0pt}IntVal\ {\isadigit{3}}{\isadigit{2}}\ {\isadigit{0}}{\isacharparenright}{\kern0pt}{\isacharparenright}{\kern0pt}\ {\isasymsqsupseteq}\ x%
\end{isabelle}%
\end{isamarkuptext}\isamarkuptrue%
}%
\Snip{AddToSub}{%
\isamarkuptrue%
\isacommand{optimization}\isamarkupfalse%
\ AddToSub{\isacharcolon}{\kern0pt}\ {\isachardoublequoteopen}{\isacharminus}{\kern0pt}x\ {\isacharplus}{\kern0pt}\ y\ {\isasymlongmapsto}\ y\ {\isacharminus}{\kern0pt}\ x{\isachardoublequoteclose}%
}%
\Snip{AddToSubObligation}{%
\isamarkuptrue%
\begin{isamarkuptext}%
\begin{isabelle}%
\ {\isadigit{1}}{\isachardot}{\kern0pt}\ trm(BinaryExpr\ BinSub\ y\ x)\ {\isacharless}{\kern0pt}\ trm(BinaryExpr\ BinAdd\ {\isacharparenleft}{\kern0pt}UnaryExpr\ UnaryNeg\ x{\isacharparenright}{\kern0pt}\ y)\isanewline
\ {\isadigit{2}}{\isachardot}{\kern0pt}\ BinaryExpr\ BinAdd\ {\isacharparenleft}{\kern0pt}UnaryExpr\ UnaryNeg\ x{\isacharparenright}{\kern0pt}\ y\ {\isasymsqsupseteq}\ BinaryExpr\ BinSub\ y\ x%
\end{isabelle}%
\end{isamarkuptext}\isamarkuptrue%
}%
\Snip{phase}{%
\isamarkuptrue%
\isacommand{phase}\isamarkupfalse%
\ AddCanonicalizations\isanewline
\ \ \isakeyword{terminating}\ trm\isanewline
\isakeyword{begin}%
\dots
\isacommand{end}\isamarkupfalse%
}%
\Snip{phase-example}{%
\isamarkuptrue%
\isacommand{phase}\isamarkupfalse%
\ Conditional\isanewline
\ \ \isakeyword{terminating}\ trm\isanewline
\isakeyword{begin}%
}%
\Snip{phase-example-1}{%
\isamarkuptrue%
\isacommand{optimization}\isamarkupfalse%
\ NegateCond{\isacharcolon}{\kern0pt}\ {\isachardoublequoteopen}{\isacharparenleft}{\kern0pt}{\isacharparenleft}{\kern0pt}{\isacharbang}{\kern0pt}c{\isacharparenright}{\kern0pt}\ {\isacharquery}{\kern0pt}\ t\ {\isacharcolon}{\kern0pt}\ f{\isacharparenright}{\kern0pt}\ {\isasymlongmapsto}\ {\isacharparenleft}{\kern0pt}c\ {\isacharquery}{\kern0pt}\ f\ {\isacharcolon}{\kern0pt}\ t{\isacharparenright}{\kern0pt}{\isachardoublequoteclose}%
}%
\Snip{phase-example-2}{%
\isamarkuptrue%
\isacommand{optimization}\isamarkupfalse%
\ TrueCond{\isacharcolon}{\kern0pt}\ {\isachardoublequoteopen}{\isacharparenleft}{\kern0pt}true\ {\isacharquery}{\kern0pt}\ t\ {\isacharcolon}{\kern0pt}\ f{\isacharparenright}{\kern0pt}\ {\isasymlongmapsto}\ t{\isachardoublequoteclose}%
}%
\Snip{phase-example-3}{%
\isamarkuptrue%
\isacommand{optimization}\isamarkupfalse%
\ FalseCond{\isacharcolon}{\kern0pt}\ {\isachardoublequoteopen}{\isacharparenleft}{\kern0pt}false\ {\isacharquery}{\kern0pt}\ t\ {\isacharcolon}{\kern0pt}\ f{\isacharparenright}{\kern0pt}\ {\isasymlongmapsto}\ f{\isachardoublequoteclose}%
}%
\Snip{phase-example-4}{%
\isamarkuptrue%
\isacommand{optimization}\isamarkupfalse%
\ BranchEqual{\isacharcolon}{\kern0pt}\ {\isachardoublequoteopen}{\isacharparenleft}{\kern0pt}c\ {\isacharquery}{\kern0pt}\ x\ {\isacharcolon}{\kern0pt}\ x{\isacharparenright}{\kern0pt}\ {\isasymlongmapsto}\ x{\isachardoublequoteclose}%
}%
\Snip{phase-example-5}{%
\isamarkuptrue%
\isacommand{optimization}\isamarkupfalse%
\ LessCond{\isacharcolon}{\kern0pt}\ {\isachardoublequoteopen}{\isacharparenleft}{\kern0pt}{\isacharparenleft}{\kern0pt}u\ {\isacharless}{\kern0pt}\ v{\isacharparenright}{\kern0pt}\ {\isacharquery}{\kern0pt}\ t\ {\isacharcolon}{\kern0pt}\ f{\isacharparenright}{\kern0pt}\ {\isasymlongmapsto}\ t\isanewline
\ \ \ \ \ \ \ \ \ \ \ \ \ \ \ \ \ \ \ when\ {\isacharparenleft}{\kern0pt}stamp{\isacharunderscore}{\kern0pt}under\ {\isacharparenleft}{\kern0pt}stamp{\isacharunderscore}{\kern0pt}expr\ u{\isacharparenright}{\kern0pt}\ {\isacharparenleft}{\kern0pt}stamp{\isacharunderscore}{\kern0pt}expr\ v{\isacharparenright}{\kern0pt}\ \isanewline
\ \ \ \ \ \ \ \ \ \ \ \ \ \ \ \ \ \ \ \ \ \ \ \ \ \ \ \ {\isasymand}\ wf{\isacharunderscore}{\kern0pt}stamp\ u\ {\isasymand}\ wf{\isacharunderscore}{\kern0pt}stamp\ v{\isacharparenright}{\kern0pt}{\isachardoublequoteclose}%
}%
\Snip{phase-example-6}{%
\isamarkuptrue%
\isacommand{optimization}\isamarkupfalse%
\ condition{\isacharunderscore}{\kern0pt}bounds{\isacharunderscore}{\kern0pt}y{\isacharcolon}{\kern0pt}\ {\isachardoublequoteopen}{\isacharparenleft}{\kern0pt}{\isacharparenleft}{\kern0pt}x\ {\isacharless}{\kern0pt}\ y{\isacharparenright}{\kern0pt}\ {\isacharquery}{\kern0pt}\ x\ {\isacharcolon}{\kern0pt}\ y{\isacharparenright}{\kern0pt}\ {\isasymlongmapsto}\ y\isanewline
\ \ \ \ \ \ \ \ \ \ \ \ \ \ \ \ \ \ \ when\ {\isacharparenleft}{\kern0pt}stamp{\isacharunderscore}{\kern0pt}under\ {\isacharparenleft}{\kern0pt}stamp{\isacharunderscore}{\kern0pt}expr\ y{\isacharparenright}{\kern0pt}\ {\isacharparenleft}{\kern0pt}stamp{\isacharunderscore}{\kern0pt}expr\ x{\isacharparenright}{\kern0pt}\ {\isasymand}\ wf{\isacharunderscore}{\kern0pt}stamp\ x\ {\isasymand}\ wf{\isacharunderscore}{\kern0pt}stamp\ y{\isacharparenright}{\kern0pt}{\isachardoublequoteclose}%
}%
\Snip{phase-example-7}{%
\isamarkuptrue%
\isacommand{end}\isamarkupfalse%
}%
\Snip{termination}{%
\isamarkuptrue%
\begin{isamarkuptext}%
\begin{isabelle}%
trm(UnaryExpr\ op\ x)\ {\isacharequal}{\kern0pt}\ trm(x)\ {\isacharplus}{\kern0pt}\ {\isadigit{2}}%
\end{isabelle}
\begin{isabelle}%
trm(BinaryExpr\ op\ x\ {\isacharparenleft}{\kern0pt}ConstantExpr\ cy{\isacharparenright}{\kern0pt})\ {\isacharequal}{\kern0pt}\ trm(x)\ {\isacharplus}{\kern0pt}\ {\isadigit{2}}%
\end{isabelle}
\begin{isabelle}%
trm(BinaryExpr\ op\ a\ b)\ {\isacharequal}{\kern0pt}\ trm(a)\ {\isacharplus}{\kern0pt}\ trm(b)\ {\isacharplus}{\kern0pt}\ {\isadigit{2}}%
\end{isabelle}
\begin{isabelle}%
trm(ConditionalExpr\ c\ t\ f)\ {\isacharequal}{\kern0pt}\ trm(c)\ {\isacharplus}{\kern0pt}\ trm(t)\ {\isacharplus}{\kern0pt}\ trm(f)\ {\isacharplus}{\kern0pt}\ {\isadigit{2}}%
\end{isabelle}
\begin{isabelle}%
trm(ConstantExpr\ c)\ {\isacharequal}{\kern0pt}\ {\isadigit{1}}%
\end{isabelle}
\begin{isabelle}%
trm(ParameterExpr\ ind\ s)\ {\isacharequal}{\kern0pt}\ {\isadigit{2}}%
\end{isabelle}
\begin{isabelle}%
trm(LeafExpr\ nid\ s)\ {\isacharequal}{\kern0pt}\ {\isadigit{2}}%
\end{isabelle}%
\end{isamarkuptext}\isamarkuptrue%
}%
\Snip{graph-representation}{%
\isamarkuptrue%
\begin{isamarkuptext}%
\textbf{typedef} IRGraph = 

\isa{{\isacharbraceleft}{\kern0pt}g\ {\isacharcolon}{\kern0pt}{\isacharcolon}{\kern0pt}\ ID\ {\isasymrightharpoonup}\ {\isacharparenleft}{\kern0pt}IRNode\ {\isasymtimes}\ Stamp{\isacharparenright}{\kern0pt}\ {\isachardot}{\kern0pt}\ finite\ {\isacharparenleft}{\kern0pt}dom\ g{\isacharparenright}{\kern0pt}{\isacharbraceright}{\kern0pt}}%
\end{isamarkuptext}\isamarkuptrue%
}%
\Snip{graph2tree}{%
\isamarkuptrue%
\begin{isamarkuptext}%
\induct{\isa{\mbox{}\inferrule{\mbox{g{\isasymllangle}n{\isasymrrangle}\ {\isacharequal}{\kern0pt}\ ConstantNode\ c}}{\mbox{g\ {\isasymturnstile}\ n\ {\isasymsimeq}\ ConstantExpr\ c}}}}{rep:constant}
\induct{\isa{\mbox{}\inferrule{\mbox{g{\isasymllangle}n{\isasymrrangle}\ {\isacharequal}{\kern0pt}\ ParameterNode\ i}\\\ \mbox{stamp\ g\ n\ {\isacharequal}{\kern0pt}\ s}}{\mbox{g\ {\isasymturnstile}\ n\ {\isasymsimeq}\ ParameterExpr\ i\ s}}}}{rep:parameter}
\induct{\isa{\mbox{}\inferrule{\mbox{g{\isasymllangle}n{\isasymrrangle}\ {\isacharequal}{\kern0pt}\ ConditionalNode\ c\ t\ f}\\\ \mbox{g\ {\isasymturnstile}\ c\ {\isasymsimeq}\ ce}\\\ \mbox{g\ {\isasymturnstile}\ t\ {\isasymsimeq}\ te}\\\ \mbox{g\ {\isasymturnstile}\ f\ {\isasymsimeq}\ fe}}{\mbox{g\ {\isasymturnstile}\ n\ {\isasymsimeq}\ ConditionalExpr\ ce\ te\ fe}}}}{rep:conditional}
\induct{\isa{\mbox{}\inferrule{\mbox{g{\isasymllangle}n{\isasymrrangle}\ {\isacharequal}{\kern0pt}\ AbsNode\ x}\\\ \mbox{g\ {\isasymturnstile}\ x\ {\isasymsimeq}\ xe}}{\mbox{g\ {\isasymturnstile}\ n\ {\isasymsimeq}\ UnaryExpr\ UnaryAbs\ xe}}}}{rep:unary}
\induct{\isa{\mbox{}\inferrule{\mbox{g{\isasymllangle}n{\isasymrrangle}\ {\isacharequal}{\kern0pt}\ SignExtendNode\ inputBits\ resultBits\ x}\\\ \mbox{g\ {\isasymturnstile}\ x\ {\isasymsimeq}\ xe}}{\mbox{g\ {\isasymturnstile}\ n\ {\isasymsimeq}\ UnaryExpr\ {\isacharparenleft}{\kern0pt}UnarySignExtend\ inputBits\ resultBits{\isacharparenright}{\kern0pt}\ xe}}}}{rep:convert}
\induct{\isa{\mbox{}\inferrule{\mbox{g{\isasymllangle}n{\isasymrrangle}\ {\isacharequal}{\kern0pt}\ AddNode\ x\ y}\\\ \mbox{g\ {\isasymturnstile}\ x\ {\isasymsimeq}\ xe}\\\ \mbox{g\ {\isasymturnstile}\ y\ {\isasymsimeq}\ ye}}{\mbox{g\ {\isasymturnstile}\ n\ {\isasymsimeq}\ BinaryExpr\ BinAdd\ xe\ ye}}}}{rep:binary}
\induct{\isa{\mbox{}\inferrule{\mbox{is{\isacharunderscore}{\kern0pt}preevaluated\ g{\isasymllangle}n{\isasymrrangle}}\\\ \mbox{stamp\ g\ n\ {\isacharequal}{\kern0pt}\ s}}{\mbox{g\ {\isasymturnstile}\ n\ {\isasymsimeq}\ LeafExpr\ n\ s}}}}{rep:leaf}
\induct{\isa{\mbox{}\inferrule{\mbox{g{\isasymllangle}n{\isasymrrangle}\ {\isacharequal}{\kern0pt}\ RefNode\ n{\isacharprime}{\kern0pt}}\\\ \mbox{g\ {\isasymturnstile}\ n{\isacharprime}{\kern0pt}\ {\isasymsimeq}\ e}}{\mbox{g\ {\isasymturnstile}\ n\ {\isasymsimeq}\ e}}}}{rep:ref}%
\end{isamarkuptext}\isamarkuptrue%
}%
\Snip{tree2graph}{%
\isamarkuptrue%
\begin{isamarkuptext}%
\induct{\isa{\mbox{}\inferrule{\mbox{find-matching\ g{\isadigit{2}}\ {\isacharparenleft}{\kern0pt}unary{\isacharunderscore}{\kern0pt}node\ op\ x{\isacharcomma}{\kern0pt}\ s{\isacharprime}{\kern0pt}{\isacharparenright}{\kern0pt}\ {\isacharequal}{\kern0pt}\ Some\ n}\\\ \mbox{g\ {\isasymoplus}\ xe\ {\isasymleadsto}\ {\isacharparenleft}{\kern0pt}g{\isadigit{2}}{\isacharcomma}{\kern0pt}\ x{\isacharparenright}{\kern0pt}}\\\ \mbox{s{\isacharprime}{\kern0pt}\ {\isacharequal}{\kern0pt}\ stamp{\isacharunderscore}{\kern0pt}unary\ op\ {\isacharparenleft}{\kern0pt}stamp\ g{\isadigit{2}}\ x{\isacharparenright}{\kern0pt}}}{\mbox{g\ {\isasymoplus}\ UnaryExpr\ op\ xe\ {\isasymleadsto}\ {\isacharparenleft}{\kern0pt}g{\isadigit{2}}{\isacharcomma}{\kern0pt}\ n{\isacharparenright}{\kern0pt}}}}}{unrep:unarysame}
\induct{\isa{\mbox{}\inferrule{\mbox{find-matching\ g\ {\isacharparenleft}{\kern0pt}ConstantNode\ c{\isacharcomma}{\kern0pt}\ stamp-from-value\ c{\isacharparenright}{\kern0pt}\ {\isacharequal}{\kern0pt}\ None}\\\ \mbox{n\ {\isacharequal}{\kern0pt}\ fresh-id\ g}\\\ \mbox{g{\isacharprime}{\kern0pt}\ {\isacharequal}{\kern0pt}\ insert\ n\ {\isacharparenleft}{\kern0pt}ConstantNode\ c{\isacharcomma}{\kern0pt}\ stamp-from-value\ c{\isacharparenright}{\kern0pt}\ g}}{\mbox{g\ {\isasymoplus}\ ConstantExpr\ c\ {\isasymleadsto}\ {\isacharparenleft}{\kern0pt}g{\isacharprime}{\kern0pt}{\isacharcomma}{\kern0pt}\ n{\isacharparenright}{\kern0pt}}}}}{unrep:constantnew}
\induct{\isa{\mbox{}\inferrule{\mbox{find-matching\ g\ {\isacharparenleft}{\kern0pt}ParameterNode\ i{\isacharcomma}{\kern0pt}\ s{\isacharparenright}{\kern0pt}\ {\isacharequal}{\kern0pt}\ None}\\\ \mbox{n\ {\isacharequal}{\kern0pt}\ fresh-id\ g}\\\ \mbox{g{\isacharprime}{\kern0pt}\ {\isacharequal}{\kern0pt}\ insert\ n\ {\isacharparenleft}{\kern0pt}ParameterNode\ i{\isacharcomma}{\kern0pt}\ s{\isacharparenright}{\kern0pt}\ g}}{\mbox{g\ {\isasymoplus}\ ParameterExpr\ i\ s\ {\isasymleadsto}\ {\isacharparenleft}{\kern0pt}g{\isacharprime}{\kern0pt}{\isacharcomma}{\kern0pt}\ n{\isacharparenright}{\kern0pt}}}}}{unrep:parameternew}
\induct{\isa{\mbox{}\inferrule{\mbox{find-matching\ g{\isadigit{2}}\ {\isacharparenleft}{\kern0pt}unary{\isacharunderscore}{\kern0pt}node\ op\ x{\isacharcomma}{\kern0pt}\ s{\isacharprime}{\kern0pt}{\isacharparenright}{\kern0pt}\ {\isacharequal}{\kern0pt}\ None}\\\ \mbox{g\ {\isasymoplus}\ xe\ {\isasymleadsto}\ {\isacharparenleft}{\kern0pt}g{\isadigit{2}}{\isacharcomma}{\kern0pt}\ x{\isacharparenright}{\kern0pt}}\\\ \mbox{s{\isacharprime}{\kern0pt}\ {\isacharequal}{\kern0pt}\ stamp{\isacharunderscore}{\kern0pt}unary\ op\ {\isacharparenleft}{\kern0pt}stamp\ g{\isadigit{2}}\ x{\isacharparenright}{\kern0pt}}\\\ \mbox{n\ {\isacharequal}{\kern0pt}\ fresh-id\ g{\isadigit{2}}}\\\ \mbox{g{\isacharprime}{\kern0pt}\ {\isacharequal}{\kern0pt}\ insert\ n\ {\isacharparenleft}{\kern0pt}unary{\isacharunderscore}{\kern0pt}node\ op\ x{\isacharcomma}{\kern0pt}\ s{\isacharprime}{\kern0pt}{\isacharparenright}{\kern0pt}\ g{\isadigit{2}}}}{\mbox{g\ {\isasymoplus}\ UnaryExpr\ op\ xe\ {\isasymleadsto}\ {\isacharparenleft}{\kern0pt}g{\isacharprime}{\kern0pt}{\isacharcomma}{\kern0pt}\ n{\isacharparenright}{\kern0pt}}}}}{unrep:unarynew}
\induct{\isa{\mbox{}\inferrule{\mbox{find-matching\ g{\isadigit{3}}\ {\isacharparenleft}{\kern0pt}bin{\isacharunderscore}{\kern0pt}node\ op\ x\ y{\isacharcomma}{\kern0pt}\ s{\isacharprime}{\kern0pt}{\isacharparenright}{\kern0pt}\ {\isacharequal}{\kern0pt}\ None}\\\ \mbox{g\ {\isasymoplus}\ xe\ {\isasymleadsto}\ {\isacharparenleft}{\kern0pt}g{\isadigit{2}}{\isacharcomma}{\kern0pt}\ x{\isacharparenright}{\kern0pt}}\\\ \mbox{g{\isadigit{2}}\ {\isasymoplus}\ ye\ {\isasymleadsto}\ {\isacharparenleft}{\kern0pt}g{\isadigit{3}}{\isacharcomma}{\kern0pt}\ y{\isacharparenright}{\kern0pt}}\\\ \mbox{s{\isacharprime}{\kern0pt}\ {\isacharequal}{\kern0pt}\ stamp{\isacharunderscore}{\kern0pt}binary\ op\ {\isacharparenleft}{\kern0pt}stamp\ g{\isadigit{3}}\ x{\isacharparenright}{\kern0pt}\ {\isacharparenleft}{\kern0pt}stamp\ g{\isadigit{3}}\ y{\isacharparenright}{\kern0pt}}\\\ \mbox{n\ {\isacharequal}{\kern0pt}\ fresh-id\ g{\isadigit{3}}}\\\ \mbox{g{\isacharprime}{\kern0pt}\ {\isacharequal}{\kern0pt}\ insert\ n\ {\isacharparenleft}{\kern0pt}bin{\isacharunderscore}{\kern0pt}node\ op\ x\ y{\isacharcomma}{\kern0pt}\ s{\isacharprime}{\kern0pt}{\isacharparenright}{\kern0pt}\ g{\isadigit{3}}}}{\mbox{g\ {\isasymoplus}\ BinaryExpr\ op\ xe\ ye\ {\isasymleadsto}\ {\isacharparenleft}{\kern0pt}g{\isacharprime}{\kern0pt}{\isacharcomma}{\kern0pt}\ n{\isacharparenright}{\kern0pt}}}}}{unrep:binarynew}
\induct{\isa{\mbox{}\inferrule{\mbox{stamp\ g\ n\ {\isacharequal}{\kern0pt}\ s}\\\ \mbox{is{\isacharunderscore}{\kern0pt}preevaluated\ g{\isasymllangle}n{\isasymrrangle}}}{\mbox{g\ {\isasymoplus}\ LeafExpr\ n\ s\ {\isasymleadsto}\ {\isacharparenleft}{\kern0pt}g{\isacharcomma}{\kern0pt}\ n{\isacharparenright}{\kern0pt}}}}}{unrep:leaf}%
\end{isamarkuptext}\isamarkuptrue%
}%
\Snip{preeval}{%
\isamarkuptrue%
\begin{isamarkuptext}%
\isa{is{\isacharunderscore}{\kern0pt}preevaluated\ {\isacharparenleft}{\kern0pt}InvokeNode\ n\ uu\ uv\ uw\ ux\ uy{\isacharparenright}{\kern0pt}\ {\isacharequal}{\kern0pt}\ True\isasep\isanewline%
is{\isacharunderscore}{\kern0pt}preevaluated\ {\isacharparenleft}{\kern0pt}InvokeWithExceptionNode\ n\ uz\ va\ vb\ vc\ vd\ ve{\isacharparenright}{\kern0pt}\ {\isacharequal}{\kern0pt}\ True\isasep\isanewline%
is{\isacharunderscore}{\kern0pt}preevaluated\ {\isacharparenleft}{\kern0pt}NewInstanceNode\ n\ vf\ vg\ vh{\isacharparenright}{\kern0pt}\ {\isacharequal}{\kern0pt}\ True\isasep\isanewline%
is{\isacharunderscore}{\kern0pt}preevaluated\ {\isacharparenleft}{\kern0pt}LoadFieldNode\ n\ vi\ vj\ vk{\isacharparenright}{\kern0pt}\ {\isacharequal}{\kern0pt}\ True\isasep\isanewline%
is{\isacharunderscore}{\kern0pt}preevaluated\ {\isacharparenleft}{\kern0pt}SignedDivNode\ n\ vl\ vm\ vn\ vo\ vp{\isacharparenright}{\kern0pt}\ {\isacharequal}{\kern0pt}\ True\isasep\isanewline%
is{\isacharunderscore}{\kern0pt}preevaluated\ {\isacharparenleft}{\kern0pt}SignedRemNode\ n\ vq\ vr\ vs\ vt\ vu{\isacharparenright}{\kern0pt}\ {\isacharequal}{\kern0pt}\ True\isasep\isanewline%
is{\isacharunderscore}{\kern0pt}preevaluated\ {\isacharparenleft}{\kern0pt}ValuePhiNode\ n\ vv\ vw{\isacharparenright}{\kern0pt}\ {\isacharequal}{\kern0pt}\ True\isasep\isanewline%
is{\isacharunderscore}{\kern0pt}preevaluated\ {\isacharparenleft}{\kern0pt}AbsNode\ v{\isacharparenright}{\kern0pt}\ {\isacharequal}{\kern0pt}\ False\isasep\isanewline%
is{\isacharunderscore}{\kern0pt}preevaluated\ {\isacharparenleft}{\kern0pt}AddNode\ v\ va{\isacharparenright}{\kern0pt}\ {\isacharequal}{\kern0pt}\ False\isasep\isanewline%
is{\isacharunderscore}{\kern0pt}preevaluated\ {\isacharparenleft}{\kern0pt}AndNode\ v\ va{\isacharparenright}{\kern0pt}\ {\isacharequal}{\kern0pt}\ False\isasep\isanewline%
is{\isacharunderscore}{\kern0pt}preevaluated\ {\isacharparenleft}{\kern0pt}BeginNode\ v{\isacharparenright}{\kern0pt}\ {\isacharequal}{\kern0pt}\ False\isasep\isanewline%
is{\isacharunderscore}{\kern0pt}preevaluated\ {\isacharparenleft}{\kern0pt}BytecodeExceptionNode\ v\ va\ vb{\isacharparenright}{\kern0pt}\ {\isacharequal}{\kern0pt}\ False\isasep\isanewline%
is{\isacharunderscore}{\kern0pt}preevaluated\ {\isacharparenleft}{\kern0pt}ConditionalNode\ v\ va\ vb{\isacharparenright}{\kern0pt}\ {\isacharequal}{\kern0pt}\ False\isasep\isanewline%
is{\isacharunderscore}{\kern0pt}preevaluated\ {\isacharparenleft}{\kern0pt}ConstantNode\ v{\isacharparenright}{\kern0pt}\ {\isacharequal}{\kern0pt}\ False\isasep\isanewline%
is{\isacharunderscore}{\kern0pt}preevaluated\ {\isacharparenleft}{\kern0pt}DynamicNewArrayNode\ v\ va\ vb\ vc\ vd{\isacharparenright}{\kern0pt}\ {\isacharequal}{\kern0pt}\ False\isasep\isanewline%
is{\isacharunderscore}{\kern0pt}preevaluated\ EndNode\ {\isacharequal}{\kern0pt}\ False\isasep\isanewline%
is{\isacharunderscore}{\kern0pt}preevaluated\ {\isacharparenleft}{\kern0pt}ExceptionObjectNode\ v\ va{\isacharparenright}{\kern0pt}\ {\isacharequal}{\kern0pt}\ False\isasep\isanewline%
is{\isacharunderscore}{\kern0pt}preevaluated\ {\isacharparenleft}{\kern0pt}FrameState\ v\ va\ vb\ vc{\isacharparenright}{\kern0pt}\ {\isacharequal}{\kern0pt}\ False\isasep\isanewline%
is{\isacharunderscore}{\kern0pt}preevaluated\ {\isacharparenleft}{\kern0pt}IfNode\ v\ va\ vb{\isacharparenright}{\kern0pt}\ {\isacharequal}{\kern0pt}\ False\isasep\isanewline%
is{\isacharunderscore}{\kern0pt}preevaluated\ {\isacharparenleft}{\kern0pt}IntegerBelowNode\ v\ va{\isacharparenright}{\kern0pt}\ {\isacharequal}{\kern0pt}\ False\isasep\isanewline%
is{\isacharunderscore}{\kern0pt}preevaluated\ {\isacharparenleft}{\kern0pt}IntegerEqualsNode\ v\ va{\isacharparenright}{\kern0pt}\ {\isacharequal}{\kern0pt}\ False\isasep\isanewline%
is{\isacharunderscore}{\kern0pt}preevaluated\ {\isacharparenleft}{\kern0pt}IntegerLessThanNode\ v\ va{\isacharparenright}{\kern0pt}\ {\isacharequal}{\kern0pt}\ False\isasep\isanewline%
is{\isacharunderscore}{\kern0pt}preevaluated\ {\isacharparenleft}{\kern0pt}IsNullNode\ v{\isacharparenright}{\kern0pt}\ {\isacharequal}{\kern0pt}\ False\isasep\isanewline%
is{\isacharunderscore}{\kern0pt}preevaluated\ {\isacharparenleft}{\kern0pt}KillingBeginNode\ v{\isacharparenright}{\kern0pt}\ {\isacharequal}{\kern0pt}\ False\isasep\isanewline%
is{\isacharunderscore}{\kern0pt}preevaluated\ {\isacharparenleft}{\kern0pt}LeftShiftNode\ v\ va{\isacharparenright}{\kern0pt}\ {\isacharequal}{\kern0pt}\ False\isasep\isanewline%
is{\isacharunderscore}{\kern0pt}preevaluated\ {\isacharparenleft}{\kern0pt}LogicNegationNode\ v{\isacharparenright}{\kern0pt}\ {\isacharequal}{\kern0pt}\ False\isasep\isanewline%
is{\isacharunderscore}{\kern0pt}preevaluated\ {\isacharparenleft}{\kern0pt}LoopBeginNode\ v\ va\ vb\ vc{\isacharparenright}{\kern0pt}\ {\isacharequal}{\kern0pt}\ False\isasep\isanewline%
is{\isacharunderscore}{\kern0pt}preevaluated\ {\isacharparenleft}{\kern0pt}LoopEndNode\ v{\isacharparenright}{\kern0pt}\ {\isacharequal}{\kern0pt}\ False\isasep\isanewline%
is{\isacharunderscore}{\kern0pt}preevaluated\ {\isacharparenleft}{\kern0pt}LoopExitNode\ v\ va\ vb{\isacharparenright}{\kern0pt}\ {\isacharequal}{\kern0pt}\ False\isasep\isanewline%
is{\isacharunderscore}{\kern0pt}preevaluated\ {\isacharparenleft}{\kern0pt}MergeNode\ v\ va\ vb{\isacharparenright}{\kern0pt}\ {\isacharequal}{\kern0pt}\ False\isasep\isanewline%
is{\isacharunderscore}{\kern0pt}preevaluated\ {\isacharparenleft}{\kern0pt}MethodCallTargetNode\ v\ va{\isacharparenright}{\kern0pt}\ {\isacharequal}{\kern0pt}\ False\isasep\isanewline%
is{\isacharunderscore}{\kern0pt}preevaluated\ {\isacharparenleft}{\kern0pt}MulNode\ v\ va{\isacharparenright}{\kern0pt}\ {\isacharequal}{\kern0pt}\ False\isasep\isanewline%
is{\isacharunderscore}{\kern0pt}preevaluated\ {\isacharparenleft}{\kern0pt}NarrowNode\ v\ va\ vb{\isacharparenright}{\kern0pt}\ {\isacharequal}{\kern0pt}\ False\isasep\isanewline%
is{\isacharunderscore}{\kern0pt}preevaluated\ {\isacharparenleft}{\kern0pt}NegateNode\ v{\isacharparenright}{\kern0pt}\ {\isacharequal}{\kern0pt}\ False\isasep\isanewline%
is{\isacharunderscore}{\kern0pt}preevaluated\ {\isacharparenleft}{\kern0pt}NewArrayNode\ v\ va\ vb{\isacharparenright}{\kern0pt}\ {\isacharequal}{\kern0pt}\ False\isasep\isanewline%
is{\isacharunderscore}{\kern0pt}preevaluated\ {\isacharparenleft}{\kern0pt}NotNode\ v{\isacharparenright}{\kern0pt}\ {\isacharequal}{\kern0pt}\ False\isasep\isanewline%
is{\isacharunderscore}{\kern0pt}preevaluated\ {\isacharparenleft}{\kern0pt}OrNode\ v\ va{\isacharparenright}{\kern0pt}\ {\isacharequal}{\kern0pt}\ False\isasep\isanewline%
is{\isacharunderscore}{\kern0pt}preevaluated\ {\isacharparenleft}{\kern0pt}ParameterNode\ v{\isacharparenright}{\kern0pt}\ {\isacharequal}{\kern0pt}\ False\isasep\isanewline%
is{\isacharunderscore}{\kern0pt}preevaluated\ {\isacharparenleft}{\kern0pt}PiNode\ v\ va{\isacharparenright}{\kern0pt}\ {\isacharequal}{\kern0pt}\ False\isasep\isanewline%
is{\isacharunderscore}{\kern0pt}preevaluated\ {\isacharparenleft}{\kern0pt}ReturnNode\ v\ va{\isacharparenright}{\kern0pt}\ {\isacharequal}{\kern0pt}\ False\isasep\isanewline%
is{\isacharunderscore}{\kern0pt}preevaluated\ {\isacharparenleft}{\kern0pt}RightShiftNode\ v\ va{\isacharparenright}{\kern0pt}\ {\isacharequal}{\kern0pt}\ False\isasep\isanewline%
is{\isacharunderscore}{\kern0pt}preevaluated\ {\isacharparenleft}{\kern0pt}ShortCircuitOrNode\ v\ va{\isacharparenright}{\kern0pt}\ {\isacharequal}{\kern0pt}\ False\isasep\isanewline%
is{\isacharunderscore}{\kern0pt}preevaluated\ {\isacharparenleft}{\kern0pt}SignExtendNode\ v\ va\ vb{\isacharparenright}{\kern0pt}\ {\isacharequal}{\kern0pt}\ False\isasep\isanewline%
is{\isacharunderscore}{\kern0pt}preevaluated\ {\isacharparenleft}{\kern0pt}StartNode\ v\ va{\isacharparenright}{\kern0pt}\ {\isacharequal}{\kern0pt}\ False\isasep\isanewline%
is{\isacharunderscore}{\kern0pt}preevaluated\ {\isacharparenleft}{\kern0pt}StoreFieldNode\ v\ va\ vb\ vc\ vd\ ve{\isacharparenright}{\kern0pt}\ {\isacharequal}{\kern0pt}\ False\isasep\isanewline%
is{\isacharunderscore}{\kern0pt}preevaluated\ {\isacharparenleft}{\kern0pt}SubNode\ v\ va{\isacharparenright}{\kern0pt}\ {\isacharequal}{\kern0pt}\ False\isasep\isanewline%
is{\isacharunderscore}{\kern0pt}preevaluated\ {\isacharparenleft}{\kern0pt}UnsignedRightShiftNode\ v\ va{\isacharparenright}{\kern0pt}\ {\isacharequal}{\kern0pt}\ False\isasep\isanewline%
is{\isacharunderscore}{\kern0pt}preevaluated\ {\isacharparenleft}{\kern0pt}UnwindNode\ v{\isacharparenright}{\kern0pt}\ {\isacharequal}{\kern0pt}\ False\isasep\isanewline%
is{\isacharunderscore}{\kern0pt}preevaluated\ {\isacharparenleft}{\kern0pt}ValueProxyNode\ v\ va{\isacharparenright}{\kern0pt}\ {\isacharequal}{\kern0pt}\ False\isasep\isanewline%
is{\isacharunderscore}{\kern0pt}preevaluated\ {\isacharparenleft}{\kern0pt}XorNode\ v\ va{\isacharparenright}{\kern0pt}\ {\isacharequal}{\kern0pt}\ False\isasep\isanewline%
is{\isacharunderscore}{\kern0pt}preevaluated\ {\isacharparenleft}{\kern0pt}ZeroExtendNode\ v\ va\ vb{\isacharparenright}{\kern0pt}\ {\isacharequal}{\kern0pt}\ False\isasep\isanewline%
is{\isacharunderscore}{\kern0pt}preevaluated\ NoNode\ {\isacharequal}{\kern0pt}\ False\isasep\isanewline%
is{\isacharunderscore}{\kern0pt}preevaluated\ {\isacharparenleft}{\kern0pt}RefNode\ v{\isacharparenright}{\kern0pt}\ {\isacharequal}{\kern0pt}\ False}%
\end{isamarkuptext}\isamarkuptrue%
}%
\Snip{deterministic-representation}{%
\isamarkuptrue%
\begin{isamarkuptext}%
\begin{isabelle}%
g\ {\isasymturnstile}\ n\ {\isasymsimeq}\ e\isactrlsub {\isadigit{1}}\ {\isasymand}\ g\ {\isasymturnstile}\ n\ {\isasymsimeq}\ e\isactrlsub {\isadigit{2}}\ {\isasymLongrightarrow}\ e\isactrlsub {\isadigit{1}}\ {\isacharequal}{\kern0pt}\ e\isactrlsub {\isadigit{2}}%
\end{isabelle}%
\end{isamarkuptext}\isamarkuptrue%
}%
\Snip{well-formed-term-graph}{%
\isamarkuptrue%
\begin{isamarkuptext}%
\isa{{\isasymexists}e{\isachardot}{\kern0pt}\ g\ {\isasymturnstile}\ n\ {\isasymsimeq}\ e\ {\isasymand}\ {\isacharparenleft}{\kern0pt}{\isasymexists}v{\isachardot}{\kern0pt}\ {\isacharbrackleft}{\kern0pt}m{\isacharcomma}{\kern0pt}p{\isacharbrackright}{\kern0pt}\ {\isasymturnstile}\ e\ {\isasymmapsto}\ v{\isacharparenright}{\kern0pt}}%
\end{isamarkuptext}\isamarkuptrue%
}%
\Snip{graph-semantics}{%
\isamarkuptrue%
\begin{isamarkuptext}%
\isa{{\isacharparenleft}{\kern0pt}{\isacharbrackleft}{\kern0pt}g{\isacharcomma}{\kern0pt}m{\isacharcomma}{\kern0pt}p{\isacharbrackright}{\kern0pt}\ {\isasymturnstile}\ n\ {\isasymmapsto}\ v{\isacharparenright}{\kern0pt}\ {\isacharequal}{\kern0pt}\ {\isacharparenleft}{\kern0pt}{\isasymexists}e{\isachardot}{\kern0pt}\ g\ {\isasymturnstile}\ n\ {\isasymsimeq}\ e\ {\isasymand}\ {\isacharbrackleft}{\kern0pt}m{\isacharcomma}{\kern0pt}p{\isacharbrackright}{\kern0pt}\ {\isasymturnstile}\ e\ {\isasymmapsto}\ v{\isacharparenright}{\kern0pt}}%
\end{isamarkuptext}\isamarkuptrue%
}%
\Snip{graph-semantics-deterministic}{%
\isamarkuptrue%
\begin{isamarkuptext}%
\isa{{\isacharbrackleft}{\kern0pt}g{\isacharcomma}{\kern0pt}m{\isacharcomma}{\kern0pt}p{\isacharbrackright}{\kern0pt}\ {\isasymturnstile}\ n\ {\isasymmapsto}\ v\isactrlsub {\isadigit{1}}\ {\isasymand}\ {\isacharbrackleft}{\kern0pt}g{\isacharcomma}{\kern0pt}m{\isacharcomma}{\kern0pt}p{\isacharbrackright}{\kern0pt}\ {\isasymturnstile}\ n\ {\isasymmapsto}\ v\isactrlsub {\isadigit{2}}\ {\isasymLongrightarrow}\ v\isactrlsub {\isadigit{1}}\ {\isacharequal}{\kern0pt}\ v\isactrlsub {\isadigit{2}}}%
\end{isamarkuptext}\isamarkuptrue%
}%
\Snip{graph-refinement}{%
\isamarkuptrue%
\begin{isamarkuptext}%
\begin{isabelle}%
term-graph-refinement\ g\isactrlsub {\isadigit{1}}\ g\isactrlsub {\isadigit{2}}\ {\isacharequal}{\kern0pt}\isanewline
{\isacharparenleft}{\kern0pt}ids\ g\isactrlsub {\isadigit{1}}\ {\isasymsubseteq}\ ids\ g\isactrlsub {\isadigit{2}}\ {\isasymand}\isanewline
\isaindent{{\isacharparenleft}{\kern0pt}}{\isacharparenleft}{\kern0pt}{\isasymforall}n{\isachardot}{\kern0pt}\ n\ {\isasymin}\ ids\ g\isactrlsub {\isadigit{1}}\ {\isasymlongrightarrow}\ {\isacharparenleft}{\kern0pt}{\isasymforall}e{\isachardot}{\kern0pt}\ g\isactrlsub {\isadigit{1}}\ {\isasymturnstile}\ n\ {\isasymsimeq}\ e\ {\isasymlongrightarrow}\ g\isactrlsub {\isadigit{2}}\ {\isasymturnstile}\ n\ {\isasymunlhd}\ e{\isacharparenright}{\kern0pt}{\isacharparenright}{\kern0pt}{\isacharparenright}{\kern0pt}%
\end{isabelle}%
\end{isamarkuptext}\isamarkuptrue%
}%
\Snip{graph-semantics-preservation}{%
\isamarkuptrue%
\begin{isamarkuptext}%
\begin{isabelle}%
e\isactrlsub {\isadigit{1}}{\isacharprime}{\kern0pt}\ {\isasymsqsupseteq}\ e\isactrlsub {\isadigit{2}}{\isacharprime}{\kern0pt}\ {\isasymand}\isanewline
{\isacharbraceleft}{\kern0pt}n{\isacharbraceright}{\kern0pt}\ $\ndres$\ g\isactrlsub {\isadigit{1}}\ {\isasymsubseteq}\ g\isactrlsub {\isadigit{2}}\ {\isasymand}\isanewline
g\isactrlsub {\isadigit{1}}\ {\isasymturnstile}\ n\ {\isasymsimeq}\ e\isactrlsub {\isadigit{1}}{\isacharprime}{\kern0pt}\ {\isasymand}\ g\isactrlsub {\isadigit{2}}\ {\isasymturnstile}\ n\ {\isasymsimeq}\ e\isactrlsub {\isadigit{2}}{\isacharprime}{\kern0pt}\ {\isasymLongrightarrow}\isanewline
term-graph-refinement\ g\isactrlsub {\isadigit{1}}\ g\isactrlsub {\isadigit{2}}%
\end{isabelle}%
\end{isamarkuptext}\isamarkuptrue%
}%
\Snip{maximal-sharing}{%
\isamarkuptrue%
\begin{isamarkuptext}%
\begin{isabelle}%
maximal{\isacharunderscore}{\kern0pt}sharing\ g\ {\isacharequal}{\kern0pt}\isanewline
{\isacharparenleft}{\kern0pt}{\isasymforall}n\isactrlsub {\isadigit{1}}\ n\isactrlsub {\isadigit{2}}{\isachardot}{\kern0pt}\isanewline
\isaindent{{\isacharparenleft}{\kern0pt}\ \ \ }n\isactrlsub {\isadigit{1}}\ {\isasymin}\ true{\isacharunderscore}{\kern0pt}ids\ g\ {\isasymand}\ n\isactrlsub {\isadigit{2}}\ {\isasymin}\ true{\isacharunderscore}{\kern0pt}ids\ g\ {\isasymlongrightarrow}\isanewline
\isaindent{{\isacharparenleft}{\kern0pt}\ \ \ }{\isacharparenleft}{\kern0pt}{\isasymforall}e{\isachardot}{\kern0pt}\ g\ {\isasymturnstile}\ n\isactrlsub {\isadigit{1}}\ {\isasymsimeq}\ e\ {\isasymand}\isanewline
\isaindent{{\isacharparenleft}{\kern0pt}\ \ \ {\isacharparenleft}{\kern0pt}{\isasymforall}e{\isachardot}{\kern0pt}\ }g\ {\isasymturnstile}\ n\isactrlsub {\isadigit{2}}\ {\isasymsimeq}\ e\ {\isasymand}\ stamp\ g\ n\isactrlsub {\isadigit{1}}\ {\isacharequal}{\kern0pt}\ stamp\ g\ n\isactrlsub {\isadigit{2}}\ {\isasymlongrightarrow}\isanewline
\isaindent{{\isacharparenleft}{\kern0pt}\ \ \ {\isacharparenleft}{\kern0pt}{\isasymforall}e{\isachardot}{\kern0pt}\ }n\isactrlsub {\isadigit{1}}\ {\isacharequal}{\kern0pt}\ n\isactrlsub {\isadigit{2}}{\isacharparenright}{\kern0pt}{\isacharparenright}{\kern0pt}%
\end{isabelle}%
\end{isamarkuptext}\isamarkuptrue%
}%
\Snip{tree-to-graph-rewriting}{%
\isamarkuptrue%
\begin{isamarkuptext}%
\begin{isabelle}%
e\isactrlsub {\isadigit{1}}\ {\isasymsqsupseteq}\ e\isactrlsub {\isadigit{2}}\ {\isasymand}\isanewline
g\isactrlsub {\isadigit{1}}\ {\isasymturnstile}\ n\ {\isasymsimeq}\ e\isactrlsub {\isadigit{1}}\ {\isasymand}\isanewline
maximal{\isacharunderscore}{\kern0pt}sharing\ g\isactrlsub {\isadigit{1}}\ {\isasymand}\isanewline
{\isacharbraceleft}{\kern0pt}n{\isacharbraceright}{\kern0pt}\ $\ndres$\ g\isactrlsub {\isadigit{1}}\ {\isasymsubseteq}\ g\isactrlsub {\isadigit{2}}\ {\isasymand}\isanewline
g\isactrlsub {\isadigit{2}}\ {\isasymturnstile}\ n\ {\isasymsimeq}\ e\isactrlsub {\isadigit{2}}\ {\isasymand}\isanewline
maximal{\isacharunderscore}{\kern0pt}sharing\ g\isactrlsub {\isadigit{2}}\ {\isasymLongrightarrow}\isanewline
term-graph-refinement\ g\isactrlsub {\isadigit{1}}\ g\isactrlsub {\isadigit{2}}%
\end{isabelle}%
\end{isamarkuptext}\isamarkuptrue%
}%
\Snip{term-graph-refines-term}{%
\isamarkuptrue%
\begin{isamarkuptext}%
\begin{isabelle}%
{\isacharparenleft}{\kern0pt}g\ {\isasymturnstile}\ n\ {\isasymunlhd}\ e{\isacharparenright}{\kern0pt}\ {\isacharequal}{\kern0pt}\ {\isacharparenleft}{\kern0pt}{\isasymexists}e{\isacharprime}{\kern0pt}{\isachardot}{\kern0pt}\ g\ {\isasymturnstile}\ n\ {\isasymsimeq}\ e{\isacharprime}{\kern0pt}\ {\isasymand}\ e\ {\isasymsqsupseteq}\ e{\isacharprime}{\kern0pt}{\isacharparenright}{\kern0pt}%
\end{isabelle}%
\end{isamarkuptext}\isamarkuptrue%
}%
\Snip{term-graph-evaluation}{%
\isamarkuptrue%
\begin{isamarkuptext}%
\begin{isabelle}%
g\ {\isasymturnstile}\ n\ {\isasymunlhd}\ e\ {\isasymLongrightarrow}\ {\isasymforall}m\ p\ v{\isachardot}{\kern0pt}\ {\isacharbrackleft}{\kern0pt}m{\isacharcomma}{\kern0pt}p{\isacharbrackright}{\kern0pt}\ {\isasymturnstile}\ e\ {\isasymmapsto}\ v\ {\isasymlongrightarrow}\ {\isacharbrackleft}{\kern0pt}g{\isacharcomma}{\kern0pt}m{\isacharcomma}{\kern0pt}p{\isacharbrackright}{\kern0pt}\ {\isasymturnstile}\ n\ {\isasymmapsto}\ v%
\end{isabelle}%
\end{isamarkuptext}\isamarkuptrue%
}%
\Snip{graph-construction}{%
\isamarkuptrue%
\begin{isamarkuptext}%
\begin{isabelle}%
e\isactrlsub {\isadigit{1}}\ {\isasymsqsupseteq}\ e\isactrlsub {\isadigit{2}}\ {\isasymand}\ g\isactrlsub {\isadigit{1}}\ {\isasymsubseteq}\ g\isactrlsub {\isadigit{2}}\ {\isasymand}\ g\isactrlsub {\isadigit{2}}\ {\isasymturnstile}\ n\ {\isasymsimeq}\ e\isactrlsub {\isadigit{2}}\ {\isasymLongrightarrow}\isanewline
g\isactrlsub {\isadigit{2}}\ {\isasymturnstile}\ n\ {\isasymunlhd}\ e\isactrlsub {\isadigit{1}}\ {\isasymand}\ term-graph-refinement\ g\isactrlsub {\isadigit{1}}\ g\isactrlsub {\isadigit{2}}%
\end{isabelle}%
\end{isamarkuptext}\isamarkuptrue%
}%
\Snip{term-graph-reconstruction}{%
\isamarkuptrue%
\begin{isamarkuptext}%
\begin{isabelle}%
g\ {\isasymoplus}\ e\ {\isasymleadsto}\ {\isacharparenleft}{\kern0pt}g{\isacharprime}{\kern0pt}{\isacharcomma}{\kern0pt}\ n{\isacharparenright}{\kern0pt}\ {\isasymLongrightarrow}\ g{\isacharprime}{\kern0pt}\ {\isasymturnstile}\ n\ {\isasymsimeq}\ e\ {\isasymand}\ g\ {\isasymsubseteq}\ g{\isacharprime}{\kern0pt}%
\end{isabelle}%
\end{isamarkuptext}\isamarkuptrue%
}%
\Snip{refined-insert}{%
\isamarkuptrue%
\begin{isamarkuptext}%
\begin{isabelle}%
e\isactrlsub {\isadigit{1}}\ {\isasymsqsupseteq}\ e\isactrlsub {\isadigit{2}}\ {\isasymand}\ g\isactrlsub {\isadigit{1}}\ {\isasymoplus}\ e\isactrlsub {\isadigit{2}}\ {\isasymleadsto}\ {\isacharparenleft}{\kern0pt}g\isactrlsub {\isadigit{2}}{\isacharcomma}{\kern0pt}\ n{\isacharprime}{\kern0pt}{\isacharparenright}{\kern0pt}\ {\isasymLongrightarrow}\isanewline
g\isactrlsub {\isadigit{2}}\ {\isasymturnstile}\ n{\isacharprime}{\kern0pt}\ {\isasymunlhd}\ e\isactrlsub {\isadigit{1}}\ {\isasymand}\ term-graph-refinement\ g\isactrlsub {\isadigit{1}}\ g\isactrlsub {\isadigit{2}}%
\end{isabelle}%
\end{isamarkuptext}\isamarkuptrue%
}%

\theoremstyle{plain}

\theoremstyle{plain}

\theoremstyle{definition}

\usepackage{laws}
\MakeEnv{lem}{Lemma}{reflem}
\MakeEnv{defi}{Definition}{refdefi}
\MakeEnv{theorem}{Theorem}{reftheorem}
\MakeEnv{corollary}{Corollary}{refcorollary}

\newcommand{\methodstate}{m}
\newcommand{\parameters}{p}
\newcommand{\RewriteOp}{\longmapsto}

\newcommand{\Rewrite}[3]{#1 \RewriteOp #2 \ifx#3\else\mathbin{\text{\textsl{when}}} #3\fi}

\newcommand{\MT}[4]{[#1,#2] \vdash #3 \mapsto #4}

\newcommand{\GT}[3]{#1 \vdash #2 \simeq #3}
\newcommand{\GTR}[3]{#1 \vdash #2 \unlhd #3}
\newcommand{\Unrep}[4]{#1 \oplus #2 \leadsto (#3, #4)}
\newcommand{\Kind}[2]{#1\langle\!\langle#2\rangle\!\rangle}

\newcommand{\dres}{\mathbin{\lhd}}
\newcommand{\ndres}{\mathbin{\rlap{\raise.05ex\hbox{$-$}}{\dres}}}
\newcommand{\fun}{\mathbin{\Rightarrow}}

\renewcommand{\implies}{\mathbin{\longrightarrow}}

\newcommand{\refsto}{\mathrel{\sqsupseteq}}

\def\draft{}

\newcommand{\figurerule}{\rule{\columnwidth}{0.5pt}}

\newcommand{\draftonly}[1]{%
\ifdefined\draft%
{\color{blue}#1}%
\else\fi}

\usepackage{ifthen}
\usepackage{calc}
\newcounter{hours}
\newcounter{minutes}
\newcommand{\printtime}{%
  \ifthenelse{\value{hours}<10}{0}{}\thehours:%
  \ifthenelse{\value{minutes}<10}{0}{}\theminutes}
\newbox{\DateTime}
\savebox{\DateTime}{\draftonly{\today\ \printtime}}
\makeatletter
\def\@setmcodes#1#2#3{{\count0=#1 \count1=#3
  \loop \global\mathcode\count0=\count1 \ifnum \count0<#2
  \advance\count0 by1 \advance\count1 by1 \repeat}}
\DeclareSymbolFont{italic}{T1}{\rmdefault}{m}{it}
\let\mathit\undefined
\DeclareSymbolFontAlphabet{\mathit}{italic}
\edef\@tempa{\hexnumber@\symitalic}
\@setmcodes{`A}{`Z}{"7\@tempa41} %
\@setmcodes{`a}{`z}{"7\@tempa61} %
\makeatother

\usepackage[colorinlistoftodos,textwidth=2cm]{todonotes}
\newcommand{\backgroundintensity}{20}
\newcommand\notesb[4]{
  \ifdefined\draft
  \todo[linecolor=red,backgroundcolor=#2!\backgroundintensity,size=\small]
  {#1: #3}{\color{blue}#4}
  \else
  \fi
}
\newcommand\notein[3]{
  \ifdefined\draft
  {\todo[inline,linecolor=red,backgroundcolor=#2!\backgroundintensity]%
  {#1: #3}
  }
  \else
  \fi
}
\newcommand\todobig[4]{\todo[inline,backgroundcolor=#2!\backgroundintensity,caption={#1 says: #3}]{ 
\begin{minipage}{\textwidth-4pt}#1: #3\par #4\end{minipage}}}

\newcommand{\notescommands}[2]{
  \expandafter\newcommand\csname #1sb\endcsname[2]{\notesb{#1}{#2}{##1}{##2}}
  \expandafter\newcommand\csname #1in\endcsname[1]{\notein{#1}{#2}{##1}}
  \expandafter\newcommand\csname #1big\endcsname[2]{\todobig{#1}{#2}{##1}{##2}}
}
\notescommands{ih}{green}
\notescommands{mu}{yellow}
\notescommands{bw}{blue}

\usepackage{alltt}

\begin{document}

\begin{abstract}
Our objective is to formally verify the correctness of the hundreds of expression
optimization rules used within the GraalVM compiler.
When defining the semantics of a programming language,
expressions naturally form abstract syntax trees, or, terms.
However, in order to facilitate sharing of common subexpressions,
modern compilers represent expressions as term graphs.
Defining the semantics of term graphs is more complicated 
than defining the semantics of their equivalent term representations.
More significantly,
defining optimizations directly on term graphs and proving semantics preservation is 
considerably more complicated than on the equivalent term representations.
On terms, optimizations can be expressed as conditional term rewriting rules,
and proofs that the rewrites are semantics preserving are relatively straightforward.
In this paper, we explore an approach to using term rewrites to verify term graph transformations
of optimizations within the GraalVM compiler.
This approach significantly reduces the overall verification effort and allows for simpler encoding of optimization rules.
\end{abstract}

\maketitle

\ifdefined\arxiv\newpage\else\fi

\section{Introduction}\labelsect{intro}

This paper focuses on verifying the optimization transformations of expressions within the GraalVM compiler \cite{graal}.
Optimizations for expressions represented as terms 
can be specified by conditional term rewriting rules.
However, in order to handle common sub-expressions,
the GraalVM compiler utilizes a term graph intermediate representation (IR) \cite{duboscq:ir:2013}
that allows common sub-expressions to be shared within the graph.
Our goal is to formally verify these expression optimizations in Isabelle/HOL \cite{IsabelleHOL}.
Our approach is to verify the corresponding term rewriting rules on the (abstract) term representation
and then give a data refinement that represents terms as term graphs 
and term rewriting rules as term graph transformations.

\vspace{-0.5em}
\paragraph{Contribution}
\begin{itemize}
    \item A framework to express and prove soundness and termination properties of optimizations expressed as term rewriting rules.
    \item A domain specific language implemented in Isabelle to express term rewriting rules using familiar Java syntax
    	and automatically generate soundness and termination proof obligations for each rule.
    \item Proofs for 45 optimization rules implemented in the compiler,
        including sophisticated bit-twiddling optimizations using static type information.
    \item A general proof that term optimizations can be lifted to the term graph optimizations used by the GraalVM compiler.
    \item Integration of proofs using a semi-automated translation from GraalVM compiler comments.
\end{itemize}
\vspace{-0.5em}

\begin{figure*}
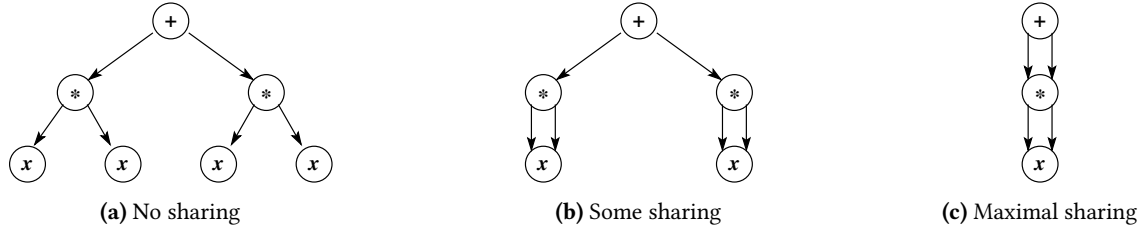

\begin{minipage}{0.4\textwidth}
\begin{center}
\input{tree1}
\subcaption{No sharing}\labelfig{tree1}
\end{center}
\end{minipage}%
\begin{minipage}{0.3\textwidth}
\begin{center}
\input{graph1}
\subcaption{Some sharing}\labelfig{graph1}
\end{center}
\end{minipage}%
\begin{minipage}{0.3\textwidth}
\begin{center}
\input{graph2}
\subcaption{Maximal sharing}\labelfig{graph2}
\end{center}
\end{minipage}
\caption{Three term graphs of the expression $(x*x)+(x*x)$ with various degrees of subgraph sharing}\labelfig{graph-sharing}
\Description{
(a) Abstract syntax tree for the term (x*x)+(x*x).
(b) Term graph of (x*x)+(x*x) such that each of the two multiplication (*) nodes share a single x node for both their arguments.
(c) Term graph of (x*x)+(x*x) with a single multiplication node acting as both inputs to the addition (+) node and a single x node for both arguments of the multiplication node.
}
\end{figure*}

\subsection{Side-Effects and Sharing in the GraalVM IR}\labelsect{side-effects}

In the GraalVM IR,
expressions are represented by 
a combined control-flow and data-flow (`sea of nodes') graph \cite{seaofnodes,duboscq:ir:2013}.
All components of an expression that may have side effects, such as method invocation nodes, 
are part of the control-flow graph. This ensures that their evaluation takes place in the correct order.
The side-effect-free components of expressions are represented by (data-flow) term sub-graphs
\cite{1979EhrigGraphGrammars,1980cStaples,1987BarendregtEtAl,2002PlumpEssentials}, %
each of which is a directed acyclic graph.
Internal nodes of a term graph represent operators, while
leaf nodes represent either constants, references to a parameter of the current method invocation,
or control-flow nodes with a pre-calculated value, 
such as a method invocation node.

For example, the expression $3 * f(x) + 1$ is represented in the IR by
a control-flow path that first visits a method invocation node, $n$, that represents invoking $f(x)$,
plus a term graph that represents the expression $3 * n + 1$.
When the node $n$ is encountered during the control-flow execution of the graph,
$f(x)$ is invoked (possibly causing side-effects), 
and the value of its result is saved in the method's state, $\methodstate$,
that is a mapping from node identifiers to values.
Any node in the subsequent control-flow traversal may evaluate the term graph corresponding to
$3 * n + 1$, in which the evaluation of the leaf node $n$ simply returns its saved value, $\methodstate(n)$.
Note that, the method invocation node $n$ plays a role in both the control-flow execution of the program
and the (side-effect-free) data-flow evaluation of any expression that includes node $n$. 
As well as method invocation nodes,
nodes corresponding to accesses to fields of objects, allocating new object instances,
and 
the equivalent of Single Static Assignment (SSA) $\phi$-nodes~\cite{ssa89} act as both control-flow nodes and 
as leaf nodes in term graphs representing side-effect-free expressions.

The use of term graphs to represent side-effect-free expressions in the GraalVM IR 
has the advantage, compared to SSA basic blocks, of not over-specifying the sequencing of evaluation of local values.
In addition, term graphs allow one to explicitly represent sharing of common sub-expressions.
For example, the expression $(x*x) + (x*x)$ can be represented by (amongst others) any of the three term graphs 
in \reffig{graph-sharing}.
Each figure represents a different degree of sharing of the common sub-expressions $x$ and $x*x$.
Because these term graphs all represent the same side-effect-free expression, 
they are semantically equivalent,
that is, they all evaluate to the same value for any given value of $x$.
However, from the point of view of generating optimized code, 
\reffig{graph2} is preferred because it maximizes sharing of common sub-expressions,
and hence it is the representation typically used in the GraalVM IR\@.
The use of a term graph with maximal sharing
replaces the use of (global) value-numbering schemes used in 
conventional SSA basic block representations, such as LLVM \cite{llvm}.
It also allows for (global) code motion optimizations,
for example, (sub-)expressions within loops, 
whose values do not change from one iteration to the next,
may be moved outside the loop and hence evaluated only once,
rather than on each iteration.

To handle an expression that may generate a runtime exception,
such as an array index out of range or a null pointer dereference,
an explicit check for the exception is inserted within the control flow graph
before the expression evaluation is reached.
Thus in optimizing expression sub-graphs, one can assume that runtime exceptions are not possible,
for example, that indices of arrays are in range and dereferenced pointers are non-null.

\subsection{Expression Semantics}\labelsect{intro-expr-semantics}

While the semantics of term graphs can be defined directly on the graph structure 
\cite{sea-of-nodes-semantics,ATVA21_GraalVM_IR_Semantics},
in doing so one needs to be concerned with well-formedness issues,
such as that the term graph contains no cycles in its evaluation and
that the term graph is \emph{closed}, that is,
any node identifiers referenced in the term graph are actually present within the graph.

From an expression evaluation semantics viewpoint 
(but not from the viewpoint of performance of generated code)
all of the term graphs in Figures~\ref{fig:tree1}--\ref{fig:graph2} are equivalent.
Unlike Figures~\ref{fig:graph1}--\ref{fig:graph2},
\reffig{tree1} also corresponds to a term or abstract syntax tree representation of the expression.
In Isabelle/HOL, the term representation of an expression can be defined by a conventional (inductive) datatype,
thus allowing the semantics of terms to be defined by a conventional inductive definition over the structure of the datatype.
Any term graph can be seen as a representation of a unique term,
for example, Figures~\ref{fig:tree1}--\ref{fig:graph2} all represent the term $(x*x)+(x*x)$,
and hence one can inductively define a relation to extract the unique term corresponding to a term graph,
thus allowing the semantics of a term graph to be defined in terms of the semantics of the corresponding term.

\subsection{Representing Expression Optimizations}\labelsect{intro-optimizations}

For side-effect-free expressions, the optimizations used in the GraalVM compiler 
are primarily based on relatively simple algebraic laws, for example, the following.
\includerawsnippet{algebraic-laws}
One can view these laws in a  number of ways:
\begin{description}
\item[on values]
as properties of values, for example, if $x$ and $y$ are fixed width integers
(of type $'a~word$ denoting a fixed bit width machine word from the Isabelle $Word$ library for any bit-width given by the type variable, $'a$)
\includerawsnippet{algebraic-laws-values}

\item[on terms]
as rewriting rules representing optimizations
\includerawsnippet{algebraic-laws-expressions}

\item[on term graphs]
as term-graph rewriting rules,\\[1ex]
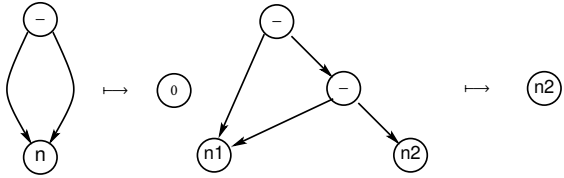
\noindent
\begin{minipage}{0.3\columnwidth}
\input{n_minus_n}
\Description{Transformation of term graph representing x-x (with node x shared) to the term graph with a single constant 0 node,
and transformation of term graph for x-(x-y) (with node x shared) to the term graph with a single y node.}
\end{minipage}%
\begin{minipage}{0.7\columnwidth}
\input{n1_minus_n1}
\end{minipage}%
\end{description}
The latter corresponds to the implementation in GraalVM\@.
An expression optimization matches a term graph with root node $r$ against the left side pattern
and, if it matches, replaces node $r$ by a new term graph for the right side,
which may contain newly created nodes (e.g.\ the node containing 0 in the first rule)
and/or reuse existing nodes (e.g.\  node $y$ in the second rule).
Proving the correctness of such an update directly on the graph representation is non-trivial because
(a) the node $r$ being updated may be referenced from multiple places within the overall graph $g$ for the method,
and
(b) a graph update has the potential to introduce cycles (although none of the optimizations updates do).

The approach that we have taken is to define each canonicalization optimization%
\footnote{Optimizations of this form are commonly referred to as ``canonicalization rules'',
even though they do not necessarily reduce semantically equivalent expressions to the same term.}
as a term rewriting rule
using a syntactic extension defined within Isabelle,
for example,

\noindent
\begin{isabelle}
\includerawsnippet{RedundantSubtract} \labelprop{RedundantSubtract}
\end{isabelle}
\raggedbottom

A rewriting rule, $\Rewrite{e_1}{e_2}{}$, is considered valid if term $e_1$ is refined by term $e_2$,
written $e_1 \refsto e_2$,
meaning that $e_2$ is well formed in any context that $e_1$ is,
and $e_1$ and $e_2$ give the same value in any such context.
For example, the proof obligation for \textsl{RedundantSubtract} is $x - (x - y) \refsto y$,
which is straightforward to prove based on the corresponding property for fixed width integer values \refprop{redundant-sub}.

The bulk of the verification work for expression optimizations is to show 
that each term rewriting rule is valid with respect to the semantics of terms,
where such proofs use the corresponding property on values.
Proofs at this more abstract level are considerably simpler than 
attempting the same proof directly on the term graph representation.
To promote these rules to the term graph representation,
we make use of a theorem that states that if a term graph at node $r$ representing a term $e_1$
is replaced by a term graph representing term $e_2$, where $e_1 \refsto e_2$,
then the original graph is refined by the updated graph.

\refsect{term-sem} gives the semantics of side-effect-free expressions on their term representations.
The semantics needs to take into account the fixed bit-width representations of integers,
for which we make extensive use of the $Word$ library in Isabelle/HOL \cite{IsabelleWordLibrary}.
\refsect{tree-trans} gives expression optimizations as term rewriting rules.
\refsect{graph-sem} defines the term graph representation for expressions,
relates that to the term representation, 
and then defines the evaluation semantics of the term graph representation via the corresponding term semantics.
\refsect{application} overviews the results of applying our approach to canonicalization optimizations 
within the GraalVM,
and \refsect{related} overviews related work.

\section{Expressions as Terms}\labelsect{term-sem}

\subsection{Representing Terms}

From this point on, \emph{expression} refers to a side-effect-free expression, the focus of this paper.
In program language semantics, an expression is represented as an abstract syntax tree or term.
For example, the expression $(x*x) + (x*x)$ is represented in our Isabelle/HOL \cite{IsabelleHOL} theory by the term, 

\begin{isabelle}
\includerawsnippet{ast-example}
\end{isabelle}
\vspace{-0.5em}
\noindent
which corresponds to the tree in \reffig{tree1}.
Terms have the advantage that they can be represented using an inductive data type,
which is simpler to work with than a term graph structure.
Terms are represented by the following Isabelle/HOL
data type, in which $IRUnaryOp$ and $IRBinaryOp$ represent
an operator (e.g.\ $BinNegate, BinAdd, \ldots$).
To increase the scope for optimization, nodes in the GraalVM IR have an associated $Stamp$
that contains typing and static analysis information about the node.
For integer values the stamp includes its bit-width and statically determined lower and upper bounds on its value.
Parameter and leaf expressions in our $IRExpr$ data type include a stamp;
the stamps for the other forms of expression can be calculated.
A leaf expression includes the node $ID$ of the corresponding control-flow graph node (e.g. a method call node).

\begin{defix}[IRExpr]
Terms are defined as an Isabelle/HOL data type.%
\footnote{The full Isabelle/HOL theories
\anon[are available as an attached artifact]{can be found at \url{https://github.com/uqcyber/veriopt-releases/tree/cpp2023}}.}
\begin{isabelle}%
\isacommand{datatype}\ IRExpr\ {\isacharequal}{\kern0pt}\isanewline
\isaindent{\ \ }UnaryExpr\ IRUnaryOp\ IRExpr\isanewline
\isaindent{\ \ }{\isacharbar}{\kern0pt}\ BinaryExpr\ IRBinaryOp\ IRExpr\ IRExpr\isanewline
\isaindent{\ \ }{\isacharbar}{\kern0pt}\ ConditionalExpr\ IRExpr\ IRExpr\ IRExpr\isanewline
\isaindent{\ \ }{\isacharbar}{\kern0pt}\ ParameterExpr\ nat\ Stamp\isanewline
\isaindent{\ \ }{\isacharbar}{\kern0pt}\ LeafExpr\ ID\ Stamp\isanewline
\isaindent{\ \ }{\isacharbar}{\kern0pt}\ ConstantExpr\ Value
\end{isabelle}%
\end{defix}

When expressions are evaluated they can return many different types of values, which we model
using the $Value$ datatype.
Integers are based on the 64-bit machine word representation using the $Word$ library of Isabelle,
which provides common arithmetic and bit-manipulation operators on fixed-width words.
Integer values additionally carry the bit width $b$, where $0 < b \leq 64$, to support integer values smaller than 64 bits.
Only the lower $b$ bits of the 64-bit word are significant; the rest are 0.

\begin{defix}[Value]{\ }
\begin{isabelle}%
\begin{isabelle}%
\isacommand{datatype}\ Value\ {\isacharequal}{\kern0pt}\isanewline
\isaindent{\ }\ UndefVal\isanewline
\isaindent{\ \ }{\isacharbar}{\kern0pt}\ IntVal\ nat\ {\isacharparenleft}{\kern0pt}{\isadigit{6}}{\isadigit{4}}\ word{\isacharparenright}{\kern0pt}\isanewline
\isaindent{\ \ }{\isacharbar}{\kern0pt}\ \dots
\end{isabelle}%

\end{isabelle}
\end{defix}

Using this $Value$ type, we define evaluation functions for unary and binary operators respectively.
For each operator, if the input values are correctly typed (for example, addition requires operands to have equal bit widths)
then the result is calculated using fixed-width operators from the $Word$ library, otherwise
$UndefVal$ is returned.
\begin{defix}[eval]{\ }
\includesnippet{eval}
\end{defix}
Using Isabelle/HOL's $Word$ library to handle fixed bit-width integers,
rather than (unbounded) mathematical integers, is essential
because it allows us to verify optimizations on the fixed-width values used in the compiler,
including the corner cases where fixed-width integers have subtly different semantics to mathematical integers. 
It also allows the verification of optimizations like 
converting a multiplication by a power of two to a left-shift operation.

\subsection{Term Semantics}

A term is evaluated with respect to the context of the method invocation in which it occurs.
That context includes a list, $\parameters$, of the values of the parameters to the method,
and the pre-calculated values of leaf expression components that may have had side effects, such as method invocation nodes.
The pre-calculated values are stored in a mapping, $\methodstate$, 
from (control flow) node identifiers to values;
we call this map the \emph{method state}.

A parameter expression is well formed if its index is within (the domain of) $\parameters$.
A leaf expression is well formed if its node identifier is in (the domain of) $\methodstate$.
Unary, binary, and conditional nodes are well formed with respect to a context $[\methodstate,\parameters]$
if their sub-expressions are well formed with respect to that context.
The relation $\MT{\methodstate}{\parameters}{e}{v}$ represents that the term $e$ is 
well formed in context $[\methodstate,\parameters]$
and evaluates to value $v$ in that context.
We use a relation in Isabelle/HOL, rather than a partial function,
because it matches a conventional (big-step) operational semantics for expressions.

Note that the evaluation relation is partial because, for example, 
one cannot evaluate a parameter with index $i$ 
if $i$ is not a valid index into the list of parameters $p$.
Evaluation assumes the term being evaluated is type correct
because the evaluation functions (e.g.\ \textit{bin-eval}) give the result $UndefVal$
if the arguments are incompatible with the operation being applied.
The notation, $v \in s$ says that a $Value$, $v$,
is contained within the range of possible values determined by the static type information ($Stamp$), $s$.

\begin{defix}[term semantics]
The semantics of a term is given in the form of a set of inference rules,
one for each syntactic form of expression.
\vspace{1ex}

\begin{minipage}{7.7cm}
\includesnippet{tree-semantics}
\end{minipage}
\end{defix}

Because evaluation is modeled as a relation, rather than a partial function, 
we need to explicitly show that evaluation is deterministic, as given by the following lemma, which has a two-line inductive proof.

\begin{lemx}[term evaluation deterministic]
For any given context of method state $m$ and parameters $p$, 
the value of a term $e$ that is well formed in that context is unique.
\includesnippet{tree-evaluation-deterministic}
\end{lemx}

\section{Domain Specific Language}\labelsect{tree-trans}

\begin{figure*}
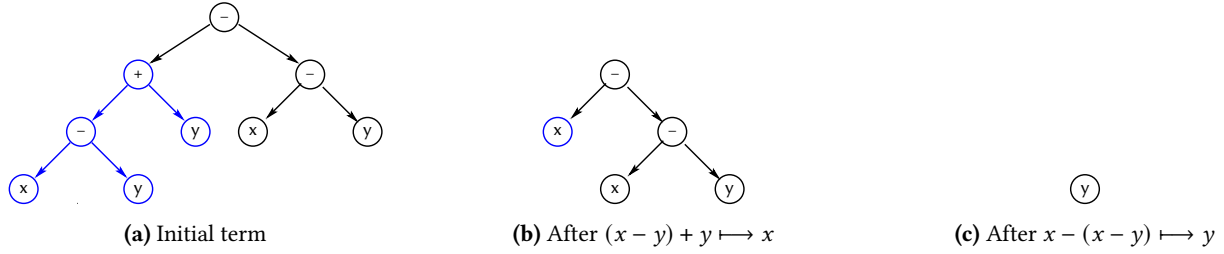

\begin{minipage}[b]{0.34\textwidth}
\begin{center}
\input{opt1}
\subcaption{Initial term}\labelfig{opt1}
\end{center}
\end{minipage}%
\begin{minipage}[b]{0.33\textwidth}
\begin{center}
\input{opt2}
\subcaption{After $\Rewrite{(x - y) + y}{x}{}$}\labelfig{opt2}
\end{center}
\end{minipage}%
\begin{minipage}[b]{0.33\textwidth}
\begin{center}
\input{opt3}
\subcaption{After $\Rewrite{x - (x - y)}{y}{}$}\labelfig{opt3}
\end{center}
\end{minipage}
\caption{Applying two rewriting rules to the tree for expression $((x-y)+y)-(x-y)$ }\labelfig{tree-rewrites}
\Description{Three abstract syntax trees (a) for ((x-y)+y)-(x-y), 
(b) for x-(x-y) after rewriting (x-y)+y to y, and
(c) for y after rewriting x-(x-y) to y.
}
\end{figure*}

To support the large number of expression optimizations in the GraalVM compiler,
we have devised a domain specific language (DSL)
that allows one to succinctly express each optimization as a term rewriting rule
and group together a set of rules into a phase.
Our terminology for term rewriting is taken  from Baader and Nipkow \cite{1998BaaderNipkow}.
To show that the set of rewriting rules in a phase is terminating, a measure function is associated with each phase.
Each optimization rule is required to:
\begin{itemize}
\item
preserve the semantics of the expression being transformed (see \refsect{preservation}),
and
\item
reduce the size of the expression according to the measure function (see \refsect{termination}).
\end{itemize}
To this end, the DSL implementation automatically generates these two proof obligations for every rule,
and to assist in the proof of these obligations, 
it automatically applies a proof tactic to simplify, and in some cases discharge, the proof obligations.

\begin{figure}[ht]
\figurerule
\begin{isabelle}
\includerawsnippet{InverseLeftSub}
\includerawsnippet{InverseLeftSubObligation}
\end{isabelle}
\figurerule
\caption{A sample optimization rule expressed as a term rewriting rule within Isabelle/HOL
and the proof obligations generated from it for (1) termination and (2) refinement.}
\labelfig{optimization-dsl-example}
\Description{Figure \ref{optimization-dsl-example} gives an optimization rule called InverseLeftSub that rewrites (x-y)+y to x.
The figure also gives the generated proof obligations:
(1) the termination condition that the ``size'' of the rewritten term is less than size of the original, and
(2) that the rewritten term, x, is a refinement of the original term, (x-y)+y.
}
\end{figure}

An optimization rule is expressed as a term rewriting rule, $\Rewrite{e_1}{e_2}{}$,
with the expression patterns written with Java-like infix notation.
\reffig{optimization-dsl-example} gives the definition of one such rule,
along with its two generated proof obligations.
The DSL is implemented within Isabelle/HOL using its syntax translation facilities \cite[Chapter 8.5]{isabelle:isar-manual}.
It makes use of a shallow embedding in Isabelle/HOL that allows one to leverage its built-in pattern matching,
enabling subexpressions in the pattern expression to be used in the transformed expression.

For a rule, $\Rewrite{e_1}{e_2}{}$,
upon a match of the pattern, $e_1$, with a term, $e$,
the pattern variables of $e_1$ will be matched with sub-terms of $e$.
Term $e$ is then replaced with, $e'$, the transformed expression consisting of
$e_2$ instantiated with the values of the pattern variables from matching pattern $e_1$ with $e$.
\reffig{tree-rewrites} shows the application of rewriting rule
$\Rewrite{(x - y) + y}{x}{}$ to the left sub-tree of \reffig{opt1} to give \reffig{opt2} and 
then rule $\Rewrite{x - (x - y)}{y}{}$ to \reffig{opt2} to give \reffig{opt3}, 
for the example expression $((x-y)+y)-(x-y)$.

\subsection{Semantics Preservation}\labelsect{preservation}

Given that term evaluation is deterministic, refinement between terms can be defined as follows.
\begin{defix}[term refinement]
A term $e_1$ is refined by another term $e_2$, written $e_1 \refsto e_2$, 
if and only if 
for all contexts $[\methodstate,\parameters]$ for which $e_1$ is well formed,
$e_2$ is well formed and evaluates to the same value as $e_1$.
\includesnippet{expression-refinement}
\end{defix}
Note that $e_2$ must be defined in all contexts in which $e_1$ is defined 
but it may be defined in more contexts than $e_1$,
e.g.\ the refinement $x-x \refsto 0$ holds but $x-x$ is only defined in contexts in which $x$ is defined, 
whereas the constant $0$ is defined in any context.
Refinement only considers whether $e_2$ is a semantically valid replacement for $e_1$,
not whether $e_2$ outperforms $e_1$ at run time.

\begin{lemx}[refinement monotone]
Refinement of any sub-term of a term gives a refinement of the whole term.

\includesnippet{expression-refinement-monotone}
\end{lemx}

\begin{figure}[ht]
\figurerule
\begin{isabelle}
\includerawsnippet{AddCommuteConstantRight}
\includerawsnippet{AddCommuteConstantRightObligation}
\end{isabelle}
\figurerule
\caption{Rule for normalizing constant expressions to the right hand side of addition operators, along with its generated proof obligations.}\labelfig{conditional-rewrite}
\Description{Figure \ref{conditional-rewrite} gives a conditional rewriting rule for swapping a constant left argument of an addition node to the right,
provided the right argument is not also a constant. 
The figure also shows the generated proof obligations:
(1) the termination condition that the ``size'' of the rewritten term is less than the original, if the condition is satisfied, and
(2) that the rewritten term is a refinement of the original, if the condition holds.
}
\end{figure}

For an unconditional term rewriting rule, $\Rewrite{e_1}{e_2}{}$, we generate the refinement obligation, $e_1 \refsto e_2$.
Conditional rewriting rules of the form, $\Rewrite{e_1}{e_2}{cond}$, restrict the rewriting to 
be applicable only to terms satisfying $cond$, and hence
the conditional term rewriting rule is valid if,
\(
  cond \implies (e_1 \refsto e_2) .
\)
For example, rule \textsl{AddCommuteConstantRight} in \reffig{conditional-rewrite} normalizes expressions
involving constants by moving constants to the right 
and has a constraint that the second argument of the plus is not already a constant.
The abbreviation $const~c$ in the optimization syntax stands for $ConstantExpr~c$.
The free pattern variables in $cond$ are instantiated with the values of the pattern variables from the match of $e_1$ with $e$.

\subsection{Optimization Phases}\labelsect{tree-systems}

A set of term rewriting rules is called a \emph{term rewriting system} or an \emph{optimization phase}.
The rules in the set can be repeatedly applied to optimize a term, and we are 
interested in various properties of the optimization phase as a whole.
Within the Isabelle/HOL representation, sets of term rewriting rules are organized into phases,
each associated with the kind of the root node of its pattern, (e.g.\ $AddNode$, $AbsNode$, $MulNode$, ...).

\begin{figure}[ht]
\figurerule
\begin{isabelle}
\includerawsnippet{phase-example}

\includerawsnippet{phase-example-1}

\includerawsnippet{phase-example-2}

\includerawsnippet{phase-example-3}

\includerawsnippet{phase-example-4}

\includerawsnippet{phase-example-5}

\includerawsnippet{phase-example-7}
\end{isabelle}
\figurerule
\caption{Illustrative subset of conditional canonicalization rules}\labelfig{conditional-rules}
\end{figure}

\reffig{conditional-rules} gives a sample optimization phase for conditional expressions. 
For the last optimization rule, \textit{LessCond},
the condition involves the stamps of expressions $u$ and $v$;
the stamps encode the results of static analyses of the expression, including range bounds,
that give upper and lower bounds on the possible values of $u$ and $v$,
allowing this optimization to be applied if the static analysis confirms that $u$ must be less than $v$
($stamp\isacharunderscore under~a~b$ is true when the upper bound of $a$ is less than the lower bound of $b$).

To optimize an expression it is sufficient to (recursively) first optimize each of the arguments of the root node
and then apply the optimization phase for the root node (i.e.\ a post-fix traversal).
Note that \reflem{refinement monotone} ensures that optimizing the arguments gives a refinement of the whole expression.

\subsection{Termination}\labelsect{termination}

A term $e$ is in \emph{normal form} (or is \emph{irreducible}) if no rewrite rules are applicable to $e$ \cite{1998BaaderNipkow}.
A set of rewriting rules is \emph{terminating} if there does not exist a term $e$ 
such that there is an infinite sequence of valid applications of the rewriting rules starting with $e$.
When devising a set of rewriting rules that will be repeatedly applied to a term,
it is important that the set of rules is terminating.
For example, a rewriting rule like
\begin{align}
  \Rewrite{x + y}{y + x}{}  \label{opt-plus-comm}
\end{align}
(while semantics preserving) can lead to an infinite sequence of rewrites,
e.g.\ $x + y \RewriteOp y + x \RewriteOp x + y \RewriteOp \cdots$.
This is undesirable because optimization phases must terminate, 
and hence this rule cannot be included in a set of optimizations.
In contrast, the rule in \reffig{conditional-rewrite} ensures termination because of its side condition.

We restrict each phase so that its set of rewriting rules must be terminating. %
Termination for a set of rewrite rules can be shown by devising a well-founded measure on terms 
(a variant function)
that strictly decreases on each application of a rewriting rule.
Phases are parameterized by a measure function, $trm$, of type, $IRExpr \fun nat$.
For each rewriting rule, $\Rewrite{e_1}{e_2}{cond}$, within a phase, 
one must show that the size of the term strictly decreases if its condition holds,
i.e.\ $cond \implies (trm~e_1 > trm~e_2)$.
This is the first proof obligation generated by the \textbf{optimization} command.
The measure below is satisfactory for the canonicalizations which have been encoded in this paper;
the measure is more complicated than usual in order to handle the asymmetric commuting rule in \reffig{conditional-rewrite}.

\begin{isabelle}
\includesnippet{termination}
\end{isabelle}

\newcommand{\measure}{trm}

Note that to handle the rule in \reffig{conditional-rewrite},
a lower weight is given to binary expressions with a constant on the right,
This ensures that
$\measure(const~v + y) = \measure(const~v) + \measure(y) + 2 = 1 + \measure(y) + 2 > 
  \measure(y) + 2 = \measure(y + const~v)$.

Phases must be parameterized by a measure as there does not exist a general measure that shows termination across all phases.
For example, multiplication by two to the power of a constant is canonicalized to left shift by the constant,
allowing more efficient generated code.
A later reassociation phase groups constants of associative operations together so that they may be constant folded.
The reassociation phase undoes the multiplication canonicalization as left shift operations are not associative and therefore limit potential reassociation.
The combined application of rules in these phases will not terminate,
instead termination relies on the sequential application of phases.

\section{Expressions as Term Graphs}\labelsect{graph-sem}

The optimization phase of the GraalVM compiler represents a side-effect-free expression as a term graph.
The main reason for using a term graph representation is to explicitly share common subexpressions
in order to generate code for a common subexpression only once.
From the point of view of the semantics of an expression, its term representation 
and all its term graph forms are equivalent
(i.e.\ evaluation in any context $[\methodstate,\parameters]$ gives the same value)
but from the point of view of generating optimized code, 
the graph with maximal sharing is preferred.
Using term graphs also improves the performance of the optimizer itself
because an optimization transformation is applied only once to a shared sub-graph,
rather than to every occurrence of the common sub-expression in a term representation
\cite{1980cStaples}.

\subsection{IR Graph Representation}\labelsect{IRNodes}

Graph nodes of the GraalVM IR are represented by the type $IRNode$,
which uses a separate node constructor for each operator corresponding to the compiler implementation, a subset of this datatype is given below where $ID$ is a numeric node identifier --- 
\anon[similar to Webb \textsl{et al.} \cite{ATVA21_GraalVM_IR_Semantics} who formalize the GraalVM IR Graph semantics]{see the Isabelle/HOL theories\footnote{\url{https://github.com/uqcyber/veriopt-releases/tree/cpp2023}} or \cite{ATVA21_GraalVM_IR_Semantics} for details}.
\begin{isabelle}
\isacommand{datatype}\ IRNode\ {\isacharequal}{\kern0pt}\isanewline
\isaindent{\ \ }AbsNode\ ID\isanewline
\isaindent{\ \ }{\isacharbar}{\kern0pt}\ AddNode\ ID\ ID\isanewline
\isaindent{\ \ }\dots\isanewline
\isaindent{\ \ }{\isacharbar}{\kern0pt}\ MulNode\ ID\ ID\isanewline
\isaindent{\ \ }{\isacharbar}{\kern0pt}\ NegateNode\ ID\isanewline
\dots\isanewline
\end{isabelle}

An IR graph, $g$, for a method is represented in Isabelle/HOL by a finite partial function from
node identifiers ($ID$) to pairs of $IRNode$ and $Stamp$;
the latter containing static type information about the node.
Finiteness is required to allow one to use induction over the nodes of the graph.
\begin{defix}[graph representation]
The graph for a method is represented by a finite mapping from node identifiers ($ID$) to nodes and their stamps.
\includesnippet{graph-representation}
\end{defix}
For a graph to be well formed any node identifiers referenced within nodes
must be in the graph (in $dom~g$) and 
the graph must not lead to cycles in evaluation of a term sub-graph.
Termination of the semantic evaluation function on the term representation is straightforward, 
but not so for graphs because they may contain cycles.%
\footnote{In the GraalVM IR, the graph for a term associated with a $\phi$-node may 
contain a cycle back to the $\phi$-node
but during evaluation of that term, the reference to the $\phi$-node 
is treated as a leaf node, thus avoiding a cycle in evaluation.}
Both these constraints are implicit in the term representation
and hence rather than explicitly constraining an $IRGraph$ to satisfy these constraints,
it is sufficient to insist that each term sub-graph represents some unique term.

\subsection{Relating Terms and Term Graphs}

For a given term $e$
there may be multiple term graph representations with different degrees of sharing of common subexpressions.
However, for a given a term graph, there is a unique corresponding term
and hence the relation between the term graph and term representations can be defined as 
a function from term graphs to terms.
The notation $\Kind{g}{n}$ stands for the $IRNode$ with id $n$ in graph $g$,
or $NoNode$ if $n$ is not in $g$,
and the notation $stamp~g~n$ extracts the stamp of the node with id $n$ within $g$.
\begin{defix}[term graph to term]
The ternary relation $\GT{g}{n}{e}$ denotes
that the term sub-graph of $g$ at $n$ represents the term $e$,
and is defined inductively by a set of inference rules, from which we give an illustrative subset below.
For rule (\ref{rep:leaf}), a node $\Kind{g}{n}$ is pre-evaluated if it is of a kind that can be used as a leaf node of a dataflow expression,
such as a method call node.

\includesnippet{graph2tree}
\end{defix}

\begin{lemx}[deterministic representation]
If the term sub-graph of $g$ at $n$ corresponds to a term $e$,
that term is unique.
\includesnippet{deterministic-representation}
\end{lemx}

If the term sub-graph of $g$ at $n$ corresponds to a term $e$,
that sub-graph must be closed 
(because for a reference to node id $n$ not in graph, $\Kind{g}{n} = NoNode$, 
which does not correspond to any term),
and not lead to a cycle in evaluation
(because the corresponding term cannot).
Furthermore, the sub-graph is well formed with respect to a context, if its corresponding term is.

\begin{defix}[term to term sub-graph]
We use the notation $\Unrep{g}{e}{g'}{n}$ to represent an operation that updates graph $g$ to give graph $g'$, 
in which node $n$ represents the term $e$.
If $e$ is already represented by some node $n$ in $g$, $g'$ is the same as $g$,
otherwise nodes are added to $g$ to give $g'$ with node $n$ being the root node of the sub-graph representing $e$.
Reusing an existing node that represents $e$, if there is one, ensures the graph preserves maximal sharing.

This relation is defined inductively by a set of inference rules, a subset of these rules are given below.
The function \textsl{find-matching} optionally returns any node that matches both the node type and stamp in the graph.
Rule \ref{unrep:unarysame} illustrates the case where the inserted node can be represented by an existing node,
the other inference rules for these cases are similar and therefore have been omitted.
Rules \ref{unrep:constantnew}-\ref{unrep:leaf} define inserting new nodes for constants, parameters, unary, binary, and leaf expressions respectively.
\end{defix}

\begin{isabelle}
\includesnippet{tree2graph}
\end{isabelle}

\begin{lemx}[term graph reconstruction]
For any insertion of a term graph representing a term, $e$, into graph, $g$,
the resulting root node identifier, $n$, in the updated graph, $g'$,
faithfully represents the original term $e$.
\includesnippet{term-graph-reconstruction}
\end{lemx}

\subsection{Term Graph Semantics}

The semantics of a term graph representation of an expression is given simply
by the semantics for its corresponding term.
\begin{defix}[term graph semantics]
Within a graph, $g$,
the semantics of a term sub-graph with root node $n$ that represents a term $e$
that is well formed for the context pair $[\methodstate,\parameters]$
is given by the semantics of the corresponding term $e$.
\includesnippet{graph-semantics}
\end{defix}

\begin{lemx}[term graph semantics deterministic]
If the semantics of a term graph is defined, it is unique.
\includesnippet{graph-semantics-deterministic}
\end{lemx}

The relation $\GT{g}{n}{e}$ states that the term graph at $n$ in $g$ (exactly) represents the term $e$,
but any term graph that represents a term $e'$ that refines $e$ is also acceptable,
and hence we define the relation $\GTR{g}{n}{e}$ that allows the term at $n$ to represent any refinement of $e$ 
(including $e$ because refinement is reflexive).
\begin{defix}[term graph refines term]
For graph $g$, node identifier $n$ and expression tree $e$,
\includesnippet{term-graph-refines-term}
\end{defix}

\begin{lemx}[term graph evaluation]
If the sub-graph of $g$ at node $n$ represents an expression that refines $e$,
evaluating the graph gives the same value as evaluating $e$ in the same context.
\includesnippet{term-graph-evaluation}
\end{lemx}

\begin{defix}[term graph refinement]
A graph $g_1$ is refined by a graph $g_2$ if,
all node identifiers in $g_1$ are also in $g_2$, and
if a node $n$ in $g_1$ represents a term $e_1$, 
it is refined by the corresponding term represented at $n$ in $g_2$.
\includesnippet{graph-refinement}
\end{defix}

A method graph $g$ also contains control-flow nodes,
which may reference term graph node ids representing expressions used by the control flow node.
\anon[]{Our earlier work gives the semantics of control-flow nodes, including method invocation \cite{ATVA21_GraalVM_IR_Semantics}.}
In all cases, the semantics of a control-flow node depends solely on 
the value of a term graph in a particular context $[m,p]$,
and not on the form of the term graph.
Hence the semantics of a control-flow node is preserved if a sub-graph representing a term $e$ it uses
is replaced by a sub-graph representing a refinement of $e$.
Detailed treatment of the semantics of control-flow nodes is given in \cite{ATVA21_GraalVM_IR_Semantics}.

\begin{figure*}
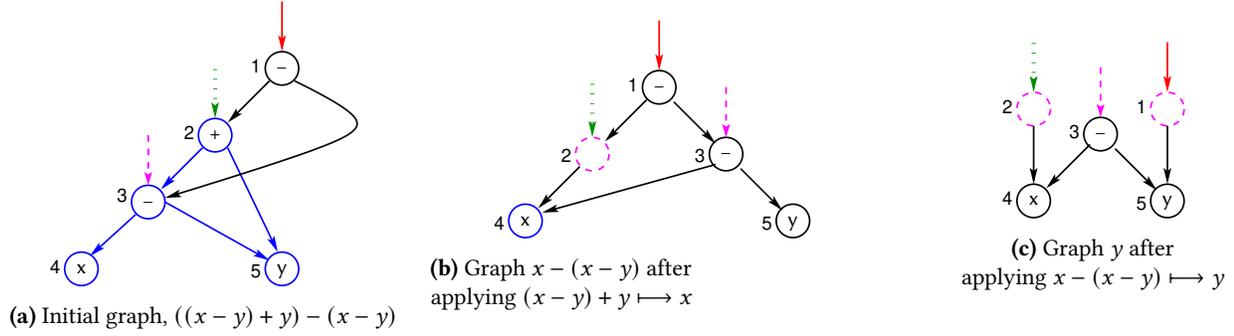

\begin{minipage}{0.34\textwidth}
\begin{center}
\input{opt1g}
\subcaption{Initial graph, $((x - y) + y) - (x -y)$}\labelfig{opt1g}
\end{center}
\end{minipage}%
\begin{minipage}{0.33\textwidth}
\begin{center}
\input{opt2r}
\subcaption{Graph $x - (x - y)$ after\\applying $\Rewrite{(x - y) + y}{x}{}$}\labelfig{opt2g}
\end{center}
\end{minipage}%
\begin{minipage}{0.33\textwidth}
\begin{center}
\input{opt3r}
\subcaption{Graph $y$ after\\applying $\Rewrite{x - (x - y)}{y}{}$}\labelfig{opt3g}
\end{center}
\end{minipage}
\caption{Maximal sharing graph for expression $((x-y)+y)-(x-y)$ with two successive rewrites.
The blue subgraph in (a) is rewritten to node 4 in (b). 
The colored solid/dashed/dotted incoming arcs to nodes 1, 2 and 3 remain unchanged throughout (despite moving their positions in the figures).
The dashed (purple) nodes in (b) and (c) are reference (indirection) nodes.}\labelfig{graph-rewrites}
\Description{Three term graphs at various stages of transformation:
(a) term graph for ((x - y) + y) - (x -y) with sharing of the subgraph for (x-y) and of the subgraph for just y,
(b) after replacing the subgraph for (x - y) + y by x to give the term graph for x - (x - y) 
but with a reference (indirection) node to the node for x as the the left argument of the root node, and
(c) after replacing the subgraph for x-(x-y) by y the original root node is now a reference node to the node for y.
}
\end{figure*}

\subsection{Term Graph Rewriting}\labelsect{term-graph-rewriting}

\reffig{graph-rewrites} gives the equivalent of the term rewriting example given in \reffig{tree-rewrites} 
for the term graph representation.
In \reffig{opt1g}, node 3, representing the term $x - y$, 
is shared as the left argument of node 2 and the right argument of node 1.
In addition,
the wider graph containing \reffig{opt1g} may have multiple references to any of the nodes in \reffig{opt1g}.
These possible references are indicated by 
the solid (red) arrow for node 1,
the dotted (green) arrow for node 2,
and the dashed (purple) arrow for node 3;
nodes 4 and 5 may also have additional references
but these have been omitted from the figure to avoid clutter.
When a term graph with root node $r$ is rewritten,
it is replaced by either a new node or some existing node $n$.
This could be handled by finding every reference to $r$ within the method graph
and replacing each with a reference to $n$ 
(as is done in the GraalVM implementation)
but with our partial mapping representation of the graph it is simpler to
replace the node at $r$ by a special indirection node, $RefNode~n$,
that redirects any reference to $r$ to node $n$.
Indirection nodes are shown as dashed (purple) circles within \reffig{graph-rewrites}.
The first rewrite from \reffig{opt1g} to \reffig{opt2g}
is applied to the term graph at node 2.
It rewrites $(x-y)+y$ to $x$ by updating node 2 to be an indirection node to node 4 ($x$).
The second rewrite from \reffig{opt2g} to \reffig{opt3g} is applied to node 1.
It rewrites $x-(x-y)$ to $y$ by replacing node 1 with an indirection node to node 5 ($y$).
Note that nodes other than $y$ remain in \reffig{opt3g}
because they may be referenced elsewhere in the larger graph.

An optimization rule, $\Rewrite{e_1}{e_2}{cond}$, succeeds at a node $n$ within $g_1$,
if the term graph at $n$ represents a (ground) term, $e_1'$, that matches $e_1$,
and $cond$ holds for that match, otherwise it fails.
If it succeeds, a term graph is constructed that represents a term $e_2'$ 
that is an instantiation of $e_2$ corresponding to the match.
The new term graph replaces the node $n$ in $g_1$ to give the new graph $g_2$.
All nodes within $g_1$ other than $n$ remain unchanged within $g_2$.
The new term graph for $e_2'$ may create new nodes (not in $g_1$) and/or
reuse existing nodes within $g_1$.
If the optimization rule is valid, (i.e.\ $cond \implies e_1 \refsto e_2$),
for a successful application of the rule that rewrites $e_1'$ to $e_2'$, it follows that $e_1' \refsto e_2'$.
The following theorem promotes term refinement to a term graph refinement between $g_1$ and $g_2$,
where for a set of node identifiers $nids$, $nids \ndres g$ stands for the graph $g$ 
with all nodes in the set $nids$ removed.
\begin{theoremx}[term graph semantics preservation]
Refining a term graph at some node $n$ within a graph, gives a refinement of the graph.
\includesnippet{graph-semantics-preservation}
\noindent
\end{theoremx}
Note that, except at $n$, the initial graph $g_1$ remains a sub-graph of the resulting graph $g_2$,
even if nodes reachable from $n$ within $g_1$ are no longer reachable from $n$ within $g_2$.
That is because such nodes may have been shared by some other expression within $g_1$ 
with a different root node, and hence may still be reachable.
If they are not used as part of some other expression in $g_1$,
they will become unreachable from the start node of $g_2$ but that does not affect the semantics.
Because the node being updated may be (indirectly) referenced from anywhere in the graph,
the proof of Theorem \ref{theorem-term graph semantics preservation} requires an induction over all possible data-flow node types,
making it quite lengthy.
\anon[In prior work on verifying GraalVM optimizations ]{In our earlier approach }\cite{ATVA21_GraalVM_IR_Semantics}, 
such a (lengthy) proof was required for each optimization rule.
Our approach \anon[]{within this paper} requires only the one proof to show that any 
optimization that is a refinement preserves the graph's semantics.

The optimization of expressions within the GraalVM compiler takes place in two ways:
\begin{enumerate}
\item\label{new-graph-constructed}
when a term graph is constructed as part of building the initial IR representation of a method,
and
\item\label{existing-graph-rewrite}
when a term graph rewriting rule is applied to an existing sub-graph to optimize it.
\end{enumerate}
In both cases a new term graph is constructed and then it is either 
(\ref{new-graph-constructed}) added to the graph, or
(\ref{existing-graph-rewrite}) replaces an (unoptimized) term graph.
\begin{theoremx}[term-graph construction]
If a term graph representing $e_1$ is required to be added to a graph, 
a term graph representing a refinement $e_2$ of $e_1$ may be added instead.
\includesnippet{graph-construction}
\end{theoremx}

\begin{corollaryx}[refined term insertion]
If a term, $e_2$, refining, $e_1$, is inserted into a graph, $g_1$, resulting in graph, $g_2$,
then the node $n'$ in $g_2$ represents $e_1$ and $g_2$ is a refinement of $g_1$.
\includesnippet{refined-insert}
\end{corollaryx}

\section{Application to GraalVM Optimizations}\labelsect{application}

We are applying this approach to the verification of the hundreds of expression optimization
rules used within the GraalVM compiler.
As such, it is beneficial to automatically discharge as many proofs as is practical.
We utilize the Eisbach \cite{Eisbach} tactic language to develop tactics that are automatically applied to any encoded optimizations.
These tactics commonly discharge optimization termination obligations and can occasionally eliminate semantic preservation obligations.
The tactics are continually evolving as common proof patterns are identified and included in the rule set.

As an example we consider the suite of optimization rules for
conditional expressions in \reffig{conditional-rules}.
In the GraalVM compiler, such a set of canonicalization rules is expressed as Java code that
matches and then rewrites the IR graph to apply the rules in a defined sequence.
\anon[Prior GraalVM verification research]{Our previous approach} \cite{ATVA21_GraalVM_IR_Semantics} involved manually translating this Java code into Isabelle IR 
graph transformation rules, and then proving directly that each rule preserves the IR graph semantics.
This involved a significant amount of tedious reasoning about the graph properties and semantics.
For example, verifying the set of rules for conditional expressions took a total of
58 lines of Isabelle proof commands.
With the more abstract term-based approach described in this paper, optimization rules can be specified
directly in Isabelle/HOL as term rewrite rules, as shown in \reffig{conditional-rules}.
Each of these rules generates two proof obligations:
\begin{enumerate}
  \item a correctness (refinement) obligation --- with automated tactics disabled,
    this takes two or three proof steps for each rule, 
    with an average of 2.4 lines/rule.  For the rules in \reffig{conditional-rules},
    this is just 12 lines of proof, 
    which is nearly five times smaller than the 58 lines required with the approach in \cite{ATVA21_GraalVM_IR_Semantics}.
    When automated tactics are enabled, all obligations without side-conditions are eliminated and 
    obligations with side-conditions are reduced to one or two proof steps.
    This results in a total of 3 lines of proof for all 5 optimizations.
  \item a termination obligation --- with automated tactics disabled,
    this is easily proven in a single line by unfolding the
    measure function then applying the \emph{simp} or \emph{auto} tactic.
    When automated tactics are enabled, all termination obligations are eliminated.
\end{enumerate}

The specifications of each optimization rule are more concise using the new approach --- 
typically one line for each rule (though in \reffig{conditional-rules} we have
added a line break before the side-conditions, for clarity), compared to an average of three lines per rule
for the term-graph approach.  As one would expect, this is also more concise than the original Java
code that implements the optimizations, which takes an average of five lines of Java code for each of these
conditional expression optimizations.

Another significant advantage is that our \anon[]{new }rewrite rules are more general.
The \anon[]{previous }graph-rewrite optimization rules \anon[ ]{\cite{ATVA21_GraalVM_IR_Semantics}} required several restrictive assumptions 
(well-formed graphs, well-formed stamps, and well formed values) that are no longer required,
because those well-formed properties can be derived from the semantics of term rewrites.
Additionally, \anon[\cite{ATVA21_GraalVM_IR_Semantics}]{previously it was} assumed that the optimized term would maintain the expression evaluation relation,
i.e.\ the proofs showed that \textsl{if} the unoptimized and optimized terms evaluated to a value \textsl{then} that value was the same.
Now the definition of term refinement ensures that the optimized term can be used in all situations where the unoptimized term was used.
And of course, our new version also proves termination, which \anon[was not addressed in \cite{ATVA21_GraalVM_IR_Semantics}]{our previous graph-rewrite approach \cite{ATVA21_GraalVM_IR_Semantics} did not address}.

In summary, our term-rewriting approach:
\begin{itemize}
  \item enables more concise specification of optimization rules,
  \item reduces the proof burden significantly,
  while proving a stronger property, and 
  \item enables termination to be proved as well.
\end{itemize}

\subsection{Workflow Integration}

Ideally, we want our verification of expression optimizations to be integrated into the development workflow of the GraalVM compiler 
so that new and modified optimizations continue to be verified.  
We do not want the compiler developers to have to learn Isabelle or become proof experts, 
but we do want a close integration between the DSL rewrite rules and the Java code that implements those rewrites.  

Our approach for verifying optimizations from the compiler is to annotate the GraalVM Java source code with inline comments that utilize our optimization DSL syntax. 
We also include a keyword (\texttt{veriopt}) to identify these comment lines, and a name for each rewrite rule.  
For example, the annotated GraalVM compiler code for most of the optimization rules in \reffig{conditional-rules} are shown in \reffig{conditional-code}.
Each return statement corresponding to a rewrite is preceded by a comment describing the rule.

\begin{figure*}[t]
\begin{lstlisting}[language=java]
if (condition instanceof LogicNegationNode) {
    LogicNegationNode negated = (LogicNegationNode) condition;
    // veriopt: NegateCond: ((!c) ? t : f) |-> (c ? f : t)
    return ConditionalNode.create(negated.getValue(), falseValue, trueValue, view);
}
if (condition instanceof LogicConstantNode) {
    LogicConstantNode c = (LogicConstantNode) condition;
    if (c.getValue()) {
        // veriopt: TrueCond: (true ? t : f) |-> t
        return trueValue;
    } else {
        // veriopt: TrueCond: (false ? t : f) |-> f
        return falseValue;
    }
}
if (trueValue == falseValue) {
    // veriopt: BranchEqual: (c ? x : x) |-> x
    return trueValue;
}
\end{lstlisting}
\caption{GraalVM compiler code to optimize a \texttt{ConditionalNode}.
The code performs most optimization rules from \reffig{conditional-rules},
the final rule is omitted for brevity.
Each optimization rule is annotated with an inline comment which explains the rule and can be used to generate proof obligations.}%
\label{fig:conditional-code}
\end{figure*}

Fortunately the GraalVM compiler already includes similar unstructured annotations to informally describe some optimizations, so this is an extension of existing practice.  
Our more precise annotations use Java-like syntax and are easily understood by developers so they are useful documentation of what transformation the Java code is performing, as well as being more concise than the corresponding Java code.
A canonicalization method in the GraalVM compiler typically handles multiple optimizations,
each corresponding to a rewriting rule.
Each optimization corresponds to a non-null return statement%
\footnote{Null returns indicate that no optimization rule for that node matches.}
that returns the rewritten term graph for the corresponding rewrite rule, 
and hence our annotation comments are placed before each such return statement.
This is a useful style guideline for the optimization phases of the compiler.

We apply an \emph{extraction tool} to the Java source code files to extract these special comment lines, and group them into Isabelle \textsl{phases}.
An appropriate termination measure is manually created or re-used for each phase.
Isabelle then generates two proof obligations for each rule to (1) ensure that the rule preserves the program semantics and (2) that the set of rules is terminating.
Tactics are applied to these obligations to attempt to automatically discharge them or else simplify the obligation.  
Any remaining proof obligations must be interactively discharged by an Isabelle proof expert.

The extraction tool also records statistics about the number of optimizations in each Java source file.  
This semi-automated process allows us to measure progress on verifying optimization rules.  
It would be useful in the future to extend the tool to check the style guideline mentioned above, 
and possibly add some fully-automated testing support that searches for counter-examples 
(similar to the Isabelle `nitpick' command) 
to the rewrite rules before they are sent to Isabelle for full verification.  
Such tooling (independent of Isabelle) will be important for these annotation comments to be accepted into the standard development lifecycle of the compiler.

\subsection{Validation}

Differential testing work as part of this project \cite{veriopt:validation}
provides confidence that the IR semantics and the specification of optimizations correspond to that of the existing GraalVM compiler.
Java programs included in the compilers existing unit tests are translated into Isabelle and executed with our executable semantics.
The output of the programs executed in Isabelle are compared with the output computed through compilation and execution with GraalVM.
Likewise, we perform optimizations of programs using GraalVM and our Isabelle specification and validate that the same optimized program is produced by both approaches.

\section{Related Work}\labelsect{related}

Representing expressions as term graphs has a long history%
\footnote{Doug McIlroy communicated privately that the earliest overt exploitation of the model that he happens to know is
Vyssotsky's use of it for Fortran data-flow diagnostics in 1962.}
\cite{DBLP:conf/stoc/AhoU70,DBLP:journals/siamcomp/AhoU72}.
The GraalVM compiler makes use of this representation as part of a larger graph combining
control-flow and data-flow as a ``sea of nodes'' \cite{seaofnodes,duboscq:ir:2013}.
Representing program transformations as conditional term rewriting rules also has a long history
\cite{IrvineCatalogue76,Maude2007,Meseguer2012}.
They have been used in language development environments for many different purposes, including
optimization transformations.  For example, the Spoofax Language Workbench for developing
domain-specific languages (DSLs) provides the \emph{Stratego} transformation language,
which extends a pure rewriting approach by introducing traversal combinators and programmable
rewriting strategies \cite{visser-rewrites}.  Stratego has been used to define many kinds 
of transformations, including optimizations and code generation.
Rewriting of term graphs \cite{2002PlumpEssentials}
has also been studied extensively, with the
TERMGRAPH conference\footnote{See \url{http://www.termgraph.org.uk}} running biannually since 2002.
The main motivation for term graph rewriting is to improve the efficiency of term rewriting
by avoiding the duplication of sub-trees.

Other research that has addressed the formal correctness of optimization rules includes the 
Alive/LLVM project \cite{Alive:2018, AliveInLean}, which uses an SMT solver to prove the correctness of peephole
optimization rules in the LLVM compiler and has found at least 64 bugs in LLVM and 13 in the SMT solver (Z3).\footnote{See \url{https://github.com/AliveToolkit/alive2} for a list of bugs found and fixed.}  
However, the peephole optimizations are defined on sequences of assembler instructions rather than graphs, so the complexities of term graph rewriting are not required.

Wider scoped compiler verification projects, such as Comp\-Cert \cite{compcert} and CakeML \cite{cakeml}, incorporate proofs of semantic preserving optimizations.
CompCert currently utilizes 10 intermediate languages,
including 2 control-flow graph based languages.
However, none of these intermediate languages explicitly represent shared sub-expressions,
instead a common sub-expression elimination phase transforms shared expressions into move instructions.
Expression optimizations similar to those presented in this paper,
are performed as peephole optimizations on sequences of assembler instructions as in Alive \cite{CompCertPeephole},
avoiding the term graph complexities.
Likewise the CakeML intermediate representation does not explicitly represent shared sub-expressions.

The VellVM project \cite{vellvm} uses the Coq theorem prover to formally verify the correctness of 
several LLVM optimizations on an SSA intermediate representation.  
Mansky and Gunter~\cite{mansky-framework-verification-optimizations}
and Strecker~\cite{STRECKER2008135} have designed generic Isabelle/HOL frameworks for proving correctness of optimizations which we anticipate to be useful for handling control-flow graph based optimizations.
The former focuses on SSA-based optimizations and have specified and proved the correctness of an SSA construction algorithm 
in their framework.
Both of those projects and VellVM performed verification directly on the
graph (or SSA) form of the program and were primarily concerned with control-flow optimizations,
whereas our focus in this paper is making it easier to verify the correctness of the multitude of expression optimization rules.

\section{Conclusions}

Our primary concern is formal proof of the correctness of each transformation rule, based
on an underlying semantics of the IR being transformed.  These proofs are more difficult when transformations
are expressed directly as term graph rewrite rules.
Our contribution is to combine the semantic simplicity of term rewriting
(leading to simplified proof obligations) with the performance-efficiency of term graph rewriting.
This gives an effective two-level approach to verifying optimizations on term graphs in Isabelle/HOL:
\begin{itemize}
  \item the first layer of verification proves the semantic correctness of each optimization rule as a simple term rewrite rule;
  \item the second layer of verification is generic, and shows that any correct tree rewrite rule can be lifted into a term graph rewrite rule that preserves the semantics of the program being optimized.
\end{itemize}
This two-layered approach is simpler than directly defining the semantics of term graphs
and directly verifying term graph transformations \cite{sea-of-nodes-semantics,ATVA21_GraalVM_IR_Semantics}
which leads to complex proofs for each rule.
Using this approach we have built a framework that allows one to more easily verify expression
canonicalization optimizations for the GraalVM IR.  

At this stage,
we have proved 45 optimization rules for the 18 currently supported expression nodes.
These rules include reasonably sophisticated bit-twiddling optimizations that rely on the static type information inferred by GraalVM.
However, they only represent a small set of the approximately 120 side-effect free compiler nodes.
The focus of the project to this stage has been on developing the proof infrastructure 
to a level where relatively inexperienced Isabelle/HOL users now can verify optimization rules
and we plan to have honours student projects working on extending the set of optimization rules handled.

Our approach is currently limited to expression optimizations that are side-effect free.
This places optimizations involving division and remainder expressions out of reach.%
\footnote{The GraalVM IR has recently been updated to include data flow versions of divide and remainder nodes,
which we do plan to handle.}
Additionally, optimizations that perform control-flow transformations are not suitable for this approach,
currently these optimizations are performed on the graph-based representation.
We plan to utilize techniques developed by Mansky \textsl{et al.} \cite{mansky-framework-verification-optimizations},
Strecker \cite{STRECKER2008135}, and Appel \cite{AppelSSA} to improve our approach to this category of optimizations.

In future work, we plan to formalize a range of \emph{strategy} operators \cite{visser-rewrites} 
for combining optimization rules in various ways, including traversing subexpressions, and 
lift our correctness and termination results to the strategy level.  We also plan to model
the generation of rewrite implementation code from rules and strategies, so that we can generate
Java code that could be integrated into the GraalVM compiler, with a clear connection back to the
verified optimization rules.

\begin{acks}
Mark Utting's position and Brae Webb's scholarship are both funded in part by a gift from Oracle Labs.
Thanks especially to Cristina Cifuentes, Paddy Krishnan, Andrew Craik and Gerg\"{o} Barany from Oracle Labs for their helpful feedback, 
and to the Oracle GraalVM compiler team for answering questions.  
Thanks also to Kristian Thomassen for his work on Isabelle/HOL proofs of canonicalization rules,
Kausthubram Rajesh for research on a domain specific language for rewriting,
and Jazmin McWilliam for work on verifying canonicalization optimizations from GraalVM using the framework of this paper.
\end{acks}

\balance
\bibliographystyle{ACM-Reference-Format}
\bibliography{references,veriopt}

\end{document}